\documentclass[11pt]{article}
\usepackage{pifont,fullpage,xspace,epsfig, wrapfig,paralist}
\usepackage{amsmath, amsthm, amssymb,times}
\usepackage{latexsym}
\usepackage{graphicx}
\usepackage{amsfonts}

\newtheorem{theorem}{Theorem}[section]
\newtheorem{lemma}[theorem]{Lemma}
\newtheorem{definition}[theorem]{Definition}

\newtheorem{proposition}[theorem]{Proposition}
\newtheorem{claim}[theorem]{Claim}

\newtheorem{corollary}[theorem]{Corollary}

\newcommand{\FI}[1]{Fig.\ \ref{#1}}
\newcommand{\ind}{\hspace*{5mm}}

\newcommand{\bec}{\leftarrow}
\newcommand{\alg}[2]{\begin{center}\fbox{\begin{minipage}{0.99\columnwidth}{\begin{center}\underline{\textsc{#1}}\end{center}{#2}}\end{minipage}}\end{center}}

\newcommand{\rbec}{\rightarrow}
\DeclareMathSymbol{\R}{\mathbin}{AMSb}{"52}

\title{Approximately Counting Embeddings into Random Graphs\thanks{A preliminary version of this paper appeared in 12th International Workshop on Randomization and Computation (RANDOM 2008).}}
\author{Martin F\"urer\thanks{Pennsylvania State University, {\tt furer@cse.psu.edu}. Research supported in part by NSF Grant CCF-0964655.} \ \ \ \ \ \ Shiva Kasiviswanathan\thanks{General Electric Research, {\tt kasivisw@gmail.com}. }}

\begin{document}
\maketitle

\begin{abstract}
Let $H$ be a graph, and let $C_H(G)$ be the number of (subgraph isomorphic) copies of $H$ contained in a graph $G$. We investigate the fundamental problem of estimating $C_H(G)$. Previous results cover only a few specific instances of this general problem, for example, the case when $H$ has degree at most one (monomer-dimer problem). In this paper, we present the first general  subcase of the subgraph isomorphism counting problem which is almost always efficiently approximable. The results rely on a new graph decomposition technique.  Informally, the decomposition is a labeling of the vertices such that every edge is between vertices with different labels and for every vertex all neighbors with a higher label have identical labels. The labeling implicitly generates a sequence of bipartite graphs which permits us to break the problem of counting embeddings of large subgraphs into that of counting embeddings of small subgraphs. Using this method, we present a simple randomized algorithm for the counting problem.  For all decomposable graphs $H$ and all graphs $G$, the algorithm is an unbiased estimator. Furthermore, for all graphs $H$ having a decomposition where each of the bipartite graphs generated is small and almost all graphs $G$, the algorithm is a fully polynomial randomized approximation scheme. 

We show that the graph classes of $H$ for which we obtain a fully polynomial randomized approximation scheme for almost all $G$ includes graphs of degree at most two, bounded-degree forests, bounded-length grid  graphs, subdivision of bounded-degree graphs,  and major subclasses of outerplanar graphs, series-parallel graphs and planar graphs, whereas unbounded-length grid graphs are excluded.  
Additionally, our general technique can easily be applied to proving many more similar results. 
\end{abstract}
\section{Introduction}
Given a {\em template} graph $H$ and a {\em base} graph $G$, we call an injection $\varphi$ between vertices of $H$ and vertices of $G$ an {\em embedding} of $H$ into $G$ if $\varphi$ maps every edge of $H$ into an edge of $G$. In other words, $\varphi$ is an isomorphism between $H$ and a subgraph (not necessarily induced) of $G$. Deciding whether such an injection exists is known as the subgraph isomorphism problem. Subgraph isomorphism is an important and general form of pattern matching. It generalizes many interesting graph problems, including Clique, Hamiltonian Path, Maximum Matching, and Shortest Path. This problem arises in application areas ranging from text processing to physics and chemistry~\cite{app1,app2,app3,app4}. The general subgraph isomorphism problem is NP-complete, but there are various special cases which are known to be fixed-parameter tractable in the size of $H$ \cite{zwick}.

In this work, we consider the related fundamental problem of counting the number of copies of a template graph in another graph. By a {\em copy} of $H$ in $G$ we mean any, not necessarily induced subgraph of $G$, isomorphic to $H$. In general the problem is \#P-complete (introduced by Valiant \cite{val}). The class \#P is defined as $\{f: \exists \mbox{ a non-deterministic polynomial time Turing machine}$ $M$ such that on input $x$, the computation tree of $M$ has exactly $f(x)$ $\mbox{accepting leaves}\}$. Problems co\-mplete for this class are presumably very difficult, especially since Toda's result \cite{toda} implies that a call to a \#P-oracle suffices to solve any problem in the polynomial hierarchy in polynomial time. 

Fixed-parameter tractability of this counting problem has been well-studied with negative results for exact counting \cite{grohe} and positive results for some special cases of approximate counting \cite{arvind}. In this paper, we are interested in the more general problem of counting copies of large subgraphs. Exact counting is possible for very few classes of non-trivial large subgraphs. A key example is perfect matchings in a planar graph~\cite{kat}. A slightly different problem that is also solvable in polynomial time is counting the number of spanning trees in a graph. A few more problems such as counting perfect matchings in a bipartite graph (a.k.a.\ $(0$-$1)$ permanent) \cite{jsv},  counting all matchings in a graph \cite{ising}, counting labeled subgraphs of a given degree sequence in a bipartite graph \cite{ivana}, counting combinatorial quantities encoded by the Tutte polynomial in a dense graph  \cite{noga1},  and counting Hamilton cycles in dense graphs \cite{fd}, can be done approximately. But problems like counting perfect matchings in general graphs are still open.

Since most of the other interesting counting problems are hopelessly hard to solve (in many cases even approximately) \cite{jer}, we investigate whether there exists a {\em fully polynomial randomized approximation scheme} (henceforth, abbreviated as fpras) that works well for {\em almost all graphs}. The statement can be made precise as: Let $G_n$ be a graph chosen uniformly at random from the set of all $n$-vertex graphs. We say that a predicate $\mathcal{P}$ holds for almost all graphs if $\Pr[\mathcal{P}(G_n)=true] \rbec 1$ as $n \rbec \infty$ (probability over the choice of a random graph). By fpras we mean a randomized algorithm  that produces a result that is correct to within a relative error of $1\pm \epsilon$ with high probability (i.e., probability tending to 1). The algorithm must run in time $\mbox{poly}(n,\epsilon^{-1})$, where $n$ is the input size. We call a problem {\em almost always efficiently approximable} if there is a randomized polynomial time algorithm producing a result within a relative error of $1\pm \epsilon$ with high probability for almost all instances.

Previous attempts at solving these kinds of problems have not been very fruitful. For example, even seemingly simple problems like counting cycles in a random graph have remained open for a long time (also stated as an open problem in the survey by Frieze and McDiarmid \cite{fmc}).  In this paper we present new techniques that can not only handle simple graphs like cycles, but also major subclasses of more complicated graph classes like outerplanar, series-parallel, planar etc.

The theory of random graphs was initiated by Erd{\H{o}}s and R{\'e}nyi \cite{erdos}. The most commonly used models of random graphs are $\mathcal{G}(n,p)$ and $\mathbb{G}(n,m)$. Both models specify a distribution on $n$-vertex graphs with a fixed set of vertices. In $\mathcal{G}(n,p)$ each of the $\binom{n}{2}$ edges is added to the graph independently with probability $p$ and $\mathbb{G}(n,m)$ assigns equal probability to all graphs with exactly $m$ edges.  Unless explicitly stated otherwise, the default model addressed in this paper is $\mathcal{G}(n,p)$. 

There has been a lot of interest in using random graph models for analyzing typical cases (beating the pessimism of worst-case analysis). Here, we mention some of these results relevant to our counting problem (see the survey of Frieze and McDiarmid \cite{fmc} for more). One of the most well-studied  problems is that of counting perfect matchings in graphs. For this problem, Jerrum and Sinclair \cite{jerrum1} have presented a simulation of a Markov chain that almost always is an fpras (extended to all bipartite graphs in \cite{jsv}). Similar results using other approaches were obtained later in~\cite{frieze,ras,chien,shiva1}. Another well-studied problem is that of counting Hamiltonian cycles in random digraphs. For this problem, Frieze and Suen \cite{suen} have obtained an fpras, and later Rasmussen \cite{ras} has presented a simpler fpras. Afterwards, Frieze \emph{et al.}\ \cite{fm} have obtained similar results in random regular graphs. Randomized approximation schemes are also available for counting the number of cliques in a random graph \cite{ras1}. However, there are no general results for counting copies of an arbitrary given~graph in a random graph. 

\subsection{Our Results and Techniques}
In this paper, we remedy this situation by presenting the first general subcase of the subgraph isomorphism counting problem that is almost always efficiently approximable. For achieving this result we introduce a new graph decomposition that we call an {\em ordered bipartite decomposition}. Informally, an ordered bipartite decomposition is a labeling of vertices such that every edge is between vertices with different labels and for every vertex all neighbors with a higher label have identical labels. The labeling implicitly generates a sequence of bipartite graphs and the crucial part is to ensure that each of the bipartite graphs is of small size. The size of the largest bipartite graph defines the {\em width} of the decomposition. The decomposition allows us to obtain general results for the counting problem which could not be achieved using the previous methods. It also leads to a relatively simple and elegant analysis. We will show that many graph classes have such a decomposition, while at the same time many simple small graphs (like a triangle) may not possess a decomposition.

The actual algorithm itself is based on the following simple sampling idea (known as importance sampling in statistics): let $\mathcal{S}=\{x_1,\dots,x_z\}$ be a large set whose cardinality we want to estimate.  Assume that we have a randomized algorithm ($\mathcal{A}$) that picks each element $x_i$ with non-zero known probability $p_i$. Then,  the Algorithm Count (\FI{count}) produces an estimate for the cardinality of $\mathcal{S}$. The following proposition shows that the estimate is unbiased, i.e., $\mathbb{E}[Z]=|\mathcal{S}|$. 

\begin{figure}[btp]
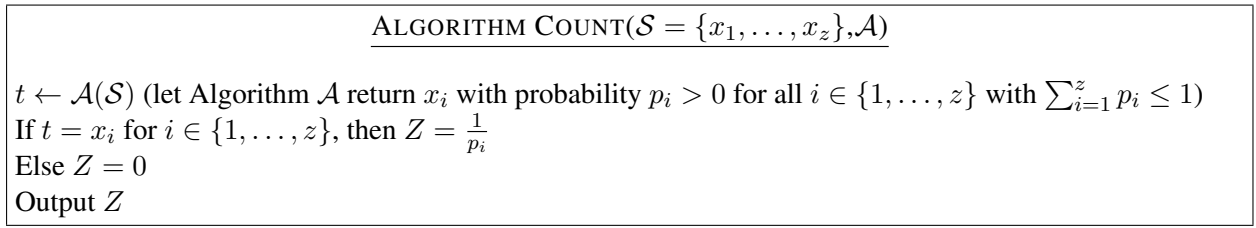

\alg{Algorithm Count($\mathcal{S} =\{x_1,\dots,x_z\}$,$\mathcal{A}$)}{
$t \leftarrow \mathcal{A}(\mathcal{S})$ (let Algorithm $\mathcal{A}$ return $x_i$ with probability $p_i > 0$ for all $i \in \{1,\dots,z\}$ with $\sum_{i=1}^{z}p_i \leq 1$)\\
If $t = x_i$ for $i \in \{1,\dots,z\}$, then $Z=\frac{1}{p_i}$ \\
Else $Z=0$ \\
Output $Z$
}
\caption{Estimator for the cardinality of $\mathcal{S}$.}
\label{count}
\end{figure}

\begin{proposition} \label{bias}
The Algorithm Count (\FI{count}) is an unbiased estimator for the cardinality of $\mathcal{S}$.
\end{proposition} 
\begin{proof}
It suffices to show that each element $x_i$ has an expected contribution of $1$ towards $|\mathcal{S}|$. This holds because on picking $x_i$ (an event that happens with probability $p_i$), we set $Z$ to the inverse probability of this event happening. Therefore, $\mathbb{E}[Z] =\sum_{i}p_i \cdot \frac{1}{p_i}=|\mathcal{S}|$.
\end{proof}

Similar schemes of counting have  previously been used by Hammersley \cite{hammer} and Knuth \cite{knuth} in other settings. Recently, this scheme has been used by Rasmussen for approximating the permanent of a $(0$-$1)$ matrix \cite{ras}, and later for approximately counting cliques in a graph \cite{ras1}. A variant of this scheme has also been used by the authors to provide a near linear-time algorithm for counting perfect matchings in random graphs \cite{shiva,shiva1}. This is however the first generalization of this simple idea to the general problem of counting graph embeddings. Another nice feature  of such schemes is that they also seem to work well in practice~\cite{piotr}.

Our randomized algorithm will try to embed $H$ into $G$. If the algorithm succeeds in finding an embedding of $H$ in $G$, it outputs the inverse probability of finding this embedding. The challenging task here is not only to ensure that each embedding of $H$ in $G$ has a positive probability of being found but also to pick each embedding with approximately equal probability to obtain a low variance. For this purpose, the algorithm considers an increasing sequence of subgraphs $\bar{H}_1 \subset \bar{H}_2 \subset \dots \subset \bar{H}_{\ell} =H$ of $H$. The algorithm starts by randomly picking an embedding of $\bar{H}_1$ into $G$, then randomly an embedding of $\bar{H}_2$ into $G$ containing the embedding of $\bar{H}_1$ and so on. It is for defining the increasing sequence of subgraphs that our decomposition is useful. 

The algorithm is always an unbiased estimator for $C_H(G)$.  The decomposition provides a natural sufficient condition for the class of algorithms based on the principle of the Algorithm Count to be an unbiased estimator. Additionally, if the base graph is a random graph from $\mathcal{G}(n,p)$ with constant $p$ and if the template graph has an ordered bipartite decomposition of bounded width, we show that the algorithm is an fpras.  The interesting case of the result is when $p=1/2$. Since the $\mathcal{G}(n,1/2)$ model assigns a uniform distribution over all graphs of $n$ given vertices, an fpras (when the base graph is from $\mathcal{G}(n,1/2)$) can be interpreted as an fpras for almost all base graphs. This result is quite powerful because now to prove that the number of copies of a template graph can be well-approximated for most graphs $G$, one just needs to show that the template graph has an ordered bipartite decomposition of bounded width.

The later half of the paper is devoted to showing that a lot of interesting graph classes naturally have an ordered bipartite decomposition of bounded width.  Let $\mathcal{C}_k$ denote a cycle of length $k$.  If a graph $H$ does not have a subgraph isomorphic to $\mathcal{C}_k$, then we say $H$ is $\mathcal{C}_k$-free.\!\footnote{This is a weaker definition that the notion of minor-free graphs used commonly in the graph theory literature~\cite{diestel}.} In this paper, we show that graphs of degree at most two, bounded-degree forests, bounded-length grid (lattice) graphs,\!\footnote{The {\em length} of an $n_1 \times n_2$ grid graph is $\min\{n_1,n_2\}$.} subdivision of bounded-degree graphs, bounded-degree outerplanar graphs which are $\mathcal{C}_3$-free, bounded-degree series-parallel graphs which are both $\mathcal{C}_3$- and $\mathcal{C}_5$-free\footnote{Denoted henceforth as $[\mathcal{C}_3,\mathcal{C}_5]$-free.}, and planar graphs of girth at least 16 have an ordered bipartite decomposition of bounded width. Using this we obtain the following result (proved in Theorems~\ref{mainproof} and~\ref{thm:probg}). 

\begin{theorem} [Main Result\footnote{The proof of this theorem follows by combining Theorems~\ref{thm:fpras} and~\ref{thm:probg}.}] \label{first}
Let $H$ be a connected graph from one of the following graph classes: graphs of degree at most two, bounded-degree trees, bounded-length grid graphs, subdivision of bounded-degree graphs, bounded-degree $\mathcal{C}_3$-free outerplanar graphs,  bounded-degree $[\mathcal{C}_3,\mathcal{C}_5]$-free series-parallel graphs, or  bounded-degree planar graphs of girth at least 16. Then,  there exists an fpras for estimating the number of copies of H in $G \in \mathcal{G}(n,p)$ for constant $p$.
\end{theorem} 

Even when restricted to graphs of degree at most two, this theorem recovers most of the older results. It also provides simpler, unified proofs for (some of) the results in \cite{frieze,ras,chien,suen}. For example, to count matchings of cardinality $k$ one could use a template consisting of $k$ disjoint edges. Similarly, to count all cycles of length $k$  the template is a cycle of that length. By varying $k$ and boosting the success probability, the algorithm can easily be extended to count all matchings or all cycles. This provides the first fpras for counting all cycles in a random graph (solving an open problem of Frieze and McDiarmid~\cite{fmc}). 

For template graphs coming from the other classes, our result supplies the first efficient randomized approximation scheme for counting copies of them in almost all base graphs. For example, it was not known earlier how to even obtain an fpras for counting the number of copies of a given bounded-degree tree in a random graph. For the simpler graph classes the decomposition follows quite straightforwardly, but for graph classes such as subdivision, outerplanar, series-parallel, and planar, constructing the decomposition requires several new combinatorial/algorithmic ideas. Even though our techniques can be extended to other interesting graph classes, we conclude by showing that our techniques can't be used to count the copies of an unbounded-length grid graph in a random graph.

\paragraph{Organization.} In Section~\ref{sec:def}, we review some useful definitions. In Section~\ref{algo}, we define the ordered bipartite decomposition, and use that to obtain an  fpras for counting copies of a graph in a random graph. Section~\ref{exam} shows that many graph classes have an ordered bipartite decomposition of bounded width, whereas in Section~\ref{neg}, we show that an unbounded-length grid graph does not have this property.  We conclude in Section~\ref{sec:concl}.

\section{Definitions and Notation} \label{sec:def}
\begin{definition} [Fully Polynomial Randomized Approximation Scheme (fpras)]
Let $Q$ be some function from the set of input strings $\Sigma^*$ to natural numbers. A fully polynomial randomized approximation scheme for $Q$ is a randomized algorithm that takes input $x \in \Sigma^*$ and an accuracy parameter $\epsilon \in (0,1)$ and outputs a number $Z$ (a random variable depending on the coin tosses of the algorithm) such that, 
\[ \Pr[(1-\epsilon)Q(x) \leq Z \leq (1+\epsilon)Q(x)] \geq 3/4, \] 
and runs in time polynomial in $|x|$, $\epsilon^{-1}$. The success probability can be boosted to $1-\delta$ by running the algorithm $O(\log \delta^{-1})$ times and taking the median \cite{vazirani}. 
\end{definition}

\paragraph{Graph Notation.}
Throughout this paper, we use $G$ to denote a base random graph on $n$ vertices.  The graph $H$ is the template whose copies we want to count in $G$. We can assume without loss of generality that the graph $H$ also contains $n$ vertices, otherwise we just add isolated vertices to $H$. The number of isomorphic images remains unaffected. Let $\triangle=\triangle(H)$ denote the maximum degree of $H$.

For a graph $F$, we use $V_F$ to denote its vertex set and $E_F$ to denote its edge set. Furthermore, we use $v_F=|V_F|$ and $e_F=|E_F|$ for the number of vertices and edges. For a subset $S$ of vertices of $F$, $N_{F}(S)=\{v \in V_F -  S\,:\,\exists u \in S \mbox{ such that }(u,v) \in E_F\}$ denotes the neighborhood of $S$ in $F$. $F[S]$ denotes the subgraph of $F$ induced by $S$. 

Automorphisms are edge respecting permutations on the set of vertices, and the set of automorphisms form a group under composition. For a graph $H$, we use $aut(H)$ to denote the size of its automorphism group. For a bounded-degree graph $H$, $aut(H)$ can be evaluated in polynomial time \cite{luks}.

We use $C_H(G)$ to denote the number of copies of $H$ in $G$. Let $L_H(G)=C_H(G) \cdot aut(H)$ denote the number of embeddings (or labeled copies) of $H$ in $G$. For a random graph $G$, we will be interested in quantities $\mathbb{E}[C_H(G)^2]$ and $\mathbb{E}[C_H(G)]^2$. 

 Most of the other graph-theoretic concepts that we use (such as planarity) are covered in standard text books (see, e.g., \cite{diestel}), and we describe them as needed. 

\paragraph{Randomization.} Our algorithm is randomized. The output of the algorithm is denoted by $Z$, which is an unbiased estimator of $C_H(G)$, i.e., $C_H(G)=\mathbb{E}_{\mathcal{A}}[Z]$ (expectation over the coin tosses of the algorithm). As the output of our algorithm depends on both the input graph, and the coin tosses of the algorithm, we use expressions such as $\mathbb{E}_{\mathcal{G}}[\mathbb{E}_{\mathcal{A}}[Z]]$. Here, the inner expectation is over the coin-tosses of the algorithm, and the outer expectation is over the graphs of $\mathcal{G} (n,p)$. Note that $\mathbb{E}_{\mathcal{A}}[Z]$ is a random variable defined on the set of graphs. 

\section{Approximation Scheme for Counting Copies} \label{algo}
We define a new graph decomposition technique which is used for embedding the template graph into the base graph.  As stated earlier our algorithm for embedding works in stages and our notion of decomposition captures this idea.   
\begin{definition} [Ordered Bipartite Decomposition]  \label{def:obd}
An ordered bipartite decomposition of a graph $H=(V_H,E_H)$ is a sequence $V_1,\dots,V_{\ell}$ of subsets of $V_H$ such that: 
\begin{dingautolist}{172}   \newcommand{\titem}{\item}
\titem $V_1,\dots,V_{\ell}$ form a partition of $V_H$.
\titem Each of the $V_i$ (for $i \in [\ell]=\{1,\dots,\ell\}$) is an independent set in $H$.   
\titem $\forall v\, \exists j$ such that $v \in V_i$ implies $N_H(v) \subseteq \left( \bigcup_{k<i}V_k \right ) \cup V_j$. 
\end{dingautolist}  
\end{definition}
Property \ding{174} just states that if a neighbor of a vertex $v \in V_i$ is in some $V_j$ ($j > i$), then all other neighbors of $v$ which are not in $V_1 \cup \dots \cup V_{i-1}$, are in $V_j$.  Property \ding{174} will be used in the analysis for random graphs to guarantee that in every stage, the base graph used for embedding is still random with the original edge probability.  

Let $V^{i} = \bigcup_{j \leq i} V_j$. Define 
\[ U_i = N_H(V_i) \cap V^{i-1}. \] 

\begin{wrapfigure}[8]{r}{100pt}
\begin{picture}(150,5)(0,80)
\put(0,-10){\epsfig{file=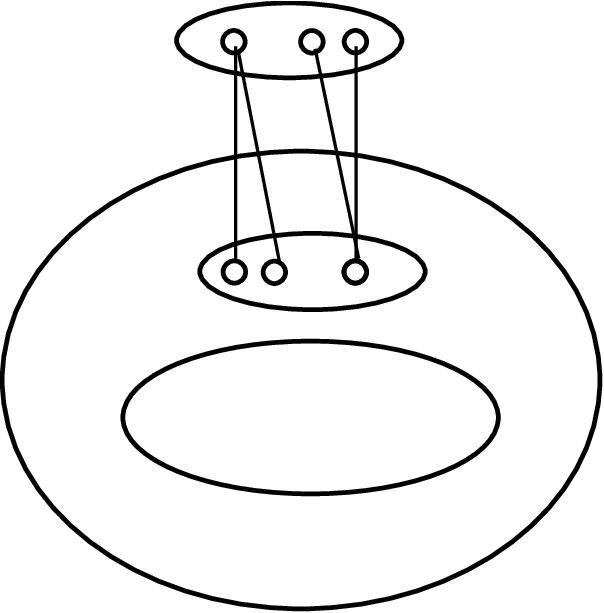,width=100pt}}
\put(44,-2){\footnotesize $V^{i-1}$}
\put(36,20){\footnotesize $\bigcup_{j < i} U_j$}
\put(72,45){\footnotesize $=U_i$}
\put(71,83){\footnotesize $=V_i$}
 \end{picture}
\end{wrapfigure} 
$U_i$ is the set of neighbors of $V_i$ in $V_1 \cup \dots \cup V_{i-1}$. Define $H_i$ to be the subgraph of $H$ induced by $U_i \cup V_i$.  Let $E_{H_i}$ denote the edge set of graph $H_i$.  
\begin{definition} [Width of Ordered Bipartite Decomposition] 
Let $V_1,\dots,V_{\ell}$ be the ordered bipartite decomposition of a graph $H=(V_H,E_H)$. Let $U_i$ be the set of neighbors of $V_i$ in $V_1 \cup \dots \cup V_{i-1}$. Define $H_i$ to be the subgraph of $H$ induced by $U_i \cup V_i$.  The {\em width} of an ordered bipartite decomposition of $H$ is the number of edges (size) in the largest $H_i$. 
\end{definition}

The $U_i$'s will play an important role in our analysis. Note that given a $U_j$, its corresponding $V_j$ has the property that $V_j \supseteq N_H(U_j)-V^{j-1}$. Hereafter, when the context is clear, we just use {\em decomposition} to denote an ordered bipartite decomposition. In general, the decomposition of a graph needn't be unique. The following lemma describes some important consequences of the decomposition. 
\begin{lemma} \label{com}
Let $V_1,\dots,V_{\ell}$ be a decomposition of a graph $H=(V_H,E_H)$. Then,  the following assertions are true. 
\begin{enumerate}
\item \label{item:1} Each of the $U_i$ is an independent set in $H$ $($$H_i$ is a bipartite graph$)$. 
\item \label{item:2} The edge set $E_H$ is partitioned into $E_{H_1},\dots,E_{H_\ell}$.
\end{enumerate}
\end{lemma}
\begin{proof}
For Part~\ref{item:1}, assume otherwise. Let $(u,v)$ be an edge in $H$ with both $u,v \in U_i$. Let $u$ appear in some $V_j$ ($j < i$) and $v$ appear in some $V_k$ ($k < i$). Property \ding{173} implies that $j \neq k$. Assume without loss of generality that $j<k$. Property \ding{174}  implies there exists no vertex $w \in N_H(u)$ such that $w \in V_i$. Therefore, $u \notin U_i$. Contradiction. Additionally, since each of the $U_i$ and $V_i$ is an independent set, each of the graph $H_i$ is bipartite.\\
For Part~\ref{item:2}, first note that due to Properties \ding{172} and \ding{174}, the $U_i$'s are pairwise disjoint (but they do not necessarily form a partition). Therefore, the $E_{H_i}$'s are also pairwise disjoint. Now since for every edge $(u,v)$ there exist a $j,k$ such that  $u \in U_j$ and $v \in V_k$ and without loss of generality $j<k$. Then, $u \in U_k$ and $(u,v) \in E_{H_k}$. Thus, $E_{H_1},\dots,E_{H_\ell}$ form a partition of $E_H$.  
\end{proof}

Every graph has a trivial decomposition satisfying Properties \ding{172} and \ding{173}, but the situation changes if we add Property \ding{174} ($\mathcal{C}_3$ is the simplest graph which has no decomposition). Every bipartite graph though has a simple decomposition, but not necessarily of bounded width. Note that the bipartiteness of $H$ is a sufficient condition for it to have an ordered bipartite decomposition, but not a necessary one.

We will primarily be interested in cases where the decomposition is of bounded width. This can only happen if $\triangle$ is a constant. In general, if $\triangle$ grows as a function of $n$, no decomposition could possibly have a bounded width ($\triangle/2$ is always a trivial lower-bound for the width). The size of the parameter $\ell$ is not important in our analysis. 
\begin{figure*}
\alg{Algorithm Embeddings(G,H)}{
Initialize $X \bec 1$, $\mathrm{\it Mark}(0) \bec \emptyset$, $\varphi(\emptyset) \bec \emptyset$ \\
Let $V_1,\dots,V_\ell$ denote an ordered bipartite decomposition of $H$ \\
For $i\bec 1$ to $\ell$ do\\
\ind Let $G_f \bec G[V_G - \mathrm{\it Mark}(i-1) \cup \varphi(U_i)]$ ($G_f$ is the subgraph of $G$ used for embedding $H_i$) \\
\ind Compute $X_i$, the number of embeddings of $H_i$ in $G_f$ with  fixed $U_i$ mapping given by~$\varphi$ \\
\ind Pick an embedding uniformly at random (if one exists) and use it to update $\varphi$\\
\ind If no embedding exists, then set $Z$ to $0$ and terminate\\
\ind $X \bec X \cdot X_i$\\
\ind $\mathrm{\it Mark}(i) \bec \mathrm{\it Mark}(i-1) \cup \varphi(V_{i})$ \\
$Z \bec X/aut(H)$ \\
Output $Z$}
\caption{Algorithms for counting copies of graph $H$ in $G$.}
\label{fig:embed}
\end{figure*}

\paragraph{Algorithm for Counting Embeddings.} The input to the Algorithm Embeddings (\FI{fig:embed}) is the template graph $H$ together with its decomposition and the base graph $G$.  The algorithm tries to construct a bijection $\varphi$ between the vertices of $H$ and $G$. $V_i$ represents the set of vertices of $H$ which get embedded into $G$ during the $i$th stage, and the already constructed mapping of $U_i$ is used to achieve this. For a subset of vertices $S \subseteq V_H$, $\varphi(S)$ denotes the image of $S$ under $\varphi$. If $X >0$ ($X$ is defined in the Algorithm Embeddings), then the function $\varphi$ represents an embedding of $H$ in $G$ (consequence of Properties \ding{172} and \ding{173}), and the output $X$ represents the inverse probability of this event happening. Since every embedding has a positive probability of being found, $X$ is an unbiased estimator for the number of embeddings of $H$ in $G$ (Proposition \ref{bias}), and $Z$ is an unbiased estimator for the number of copies of $H$ in $G$. 

The actual procedure for computing the $X_i$'s is not very relevant for our results, but note that the $X_i$'s can be computed in polynomial time if $H$ has a decomposition of bounded width.  In this case the Algorithm Embeddings runs in polynomial time.

Since the Algorithm Embeddings is an unbiased estimator, use of Chebychev's inequality implies that repeating the algorithm $O(\epsilon^{-2}\mathbb{E}_{\mathcal{A}}[Z^2]/\mathbb{E}_{\mathcal{A}}[Z]^2)$ times and taking the mean of the outputs results in a randomized approximation scheme for estimating $C_H(G)$.  The ratio $\mathbb{E}_{\mathcal{A}}[Z^2]/\mathbb{E}_{\mathcal{A}}[Z]^2$ is commonly referred to as the {\em critical ratio}.

\subsection{FPRAS for Counting in Random Graphs}
We now concentrate on showing that for random graphs the algorithm is an fpras. From here on, we abbreviate $C_H(G)$ as $C$. A few of the technical details in our proof are somewhat similar to previous applications of this sampling idea, such as that for counting perfect matchings \cite{ras,shiva1}. The simpler techniques in these previous results, however, are limited to handling one edge per stage (therefore, they work only when $H$ is a matching).  Algorithm Embeddings embeds a small sized subgraph at every stage. The key for obtaining an fpras is to guarantee that the factor contributed to the critical ratio at every stage is very small (which is now involved because it is no longer a simple ratio of binomial moments as in \cite{ras,shiva1}). We then do a stage-by-stage analysis of the critical ratio 
to show that the Algorithm Embedding is an fpras.

The analysis will be done for a worst-case graph $H$ under the assumption that the width of the decomposition of $H$ is bounded by a universal constant $w$. Here, instead of investigating the critical ratio, we investigate the much simpler ratio $\mathbb{E}_{\mathcal{G}}[\mathbb{E}_{\mathcal{A}}[Z^2]]/\mathbb{E}_{\mathcal{G}}[\mathbb{E}_{\mathcal{A}}[Z]]^2$, which we call the {\em critical ratio of averages}. We use the second moment method to show that these two ratios (critical ratio and critical ratio of averages) are closely related. To establish this fact, we take a detour through the $\mathbb{G}(n,m)$ model. The ratio $\mathbb{E}[C^2]/\mathbb{E}[C]^2$ plays an important role here and for bounding it we use a recent result of Riordan \cite{riordan}. The result (stated below) studies the related question of when a random graph $G$ is likely to have a spanning subgraph isomorphic to $H$.  Let $\triangle=\triangle(H)$ denote the maximum degree of $H$. The idea behind the following theorem is to use Markov's inequality to bound $\Pr[C=0]$ in terms of $\mathbb{E}[C]$ and $Var[C]$. The main thrust lies in proving that $\mathbb{E}[C^2]/\mathbb{E}[C]^2 = 1+o(1)$.\!\footnote{
Since $C$ is fairly tightly concentrated around its mean, a rudimentary approximation for $C$ is just $\mathbb{E}[C]=\frac{n!p^{e_H}}{aut(H)}$ (as $v_H=n$). However, this naive approach doesn't produce for {\em any} $\epsilon > 0$, an $(1\pm\epsilon)$-approximation for $C$ (see, e.g.,~\cite{frieze,suen,ras,ras1,chien}).}

In the following, $N$ is used to denote $\binom{n}{2}$. We say an event holds with high probability (w.h.p.), if it holds with probability tending to $1$ as $n \rbec \infty$.
\begin{theorem}[Riordan \cite{riordan}, Restated] \label{rior}
Let $H$ be a graph on $n$ vertices. Let $e_H=\alpha N=\alpha(n) N$, and let $p=p(n) \in (0,1)$ with $pN$ an integer. Suppose that the following conditions hold: $\alpha N \geq n$, and  $pN, (1-p)\sqrt{n}, np^{\gamma}/\triangle^4 \rbec \infty$, where  
\[\gamma=\gamma(H)=\max_{3 \leq s \leq n}\{ \max \{e_F\,:\,F \subseteq H, v_F=s\}/(s-2)\}. \]
Then,  w.h.p.\ a random graph $G \in \mathbb{G}(n,pN)$ has a spanning subgraph isomorphic to $H$.  In particular, $C=C_H(G)$ satisfies
$$\frac{\mathbb{E}[C^2]}{\mathbb{E}[C]^2} = 1+o(1).$$
\end{theorem}
The quantity $\gamma$ is closely related to twice the maximum average degree of a subgraph of $H$. 

%

The templates graph that we will be interested are bounded-degree connected graphs. For a bounded-degree graph $H$, both $\triangle$ and $\gamma$ are constants. Also, since the graph is connected $\alpha N \geq n$. Additionally, for us $p$ is a constant (as we work with dense random graphs $G$). Therefore, the conditions of Theorem~\ref{rior} are all satisfied.

\begin{corollary} \label{cor:gnm}
Let $H$ be a bounded-degree connected graph on $n$ vertices. Then,  w.h.p.\ a random graph $G \in \mathbb{G}(n,\Omega(n^2))$ satisfies $\mathbb{E}[C^2]/\mathbb{E}[C]^2 = 1+o(1)$.
\end{corollary} 

Corollary~\ref{cor:gnm} with Chebychev's inequality gives, $\Pr[C < \mathbb{E}(C)]$ tends to $0$ as $n$ tends to $\infty$. Using this and standard results on asymptotic equivalence between $\mathbb{G}(n,m)$ and $\mathcal{G}(n,p)$ models of random graphs (e.g., see Proposition 1.12 of \cite{jansonr}) yields the following corollary. Similar analysis has been used in the previous works of~\cite{frieze,suen,ras,ras1,chien}.
\begin{corollary} \label{cor:l1}
Let $H$ be a bounded-degree connected graph on $n$ vertices. Let $\omega=\omega(n)$ be any function tending to $\infty$ as $n \rbec \infty$, and let $p$ be a constant. Then,  w.h.p.\ a random graph $G \in \mathcal{G}(n,p)$ satisfies $C \geq \mathbb{E}[C]/\omega$.
\end{corollary}

Using the above result we investigate the performance of Algorithm Embeddings when $G$ is a random graph.
The proof idea is to break the critical ratio analysis of the large subgraph into a more manageable critical ratio analysis of small subgraphs.

\begin{proposition} \label{critical} \label{mainproof}
Let $H$ be an $n$-vertex connected graph with a decomposition of width $w$ (a constant). Let $Z$ be the output of Algorithm Embeddings, and let $p$ be a constant. Then,  w.h.p.\ for a random graph $G \in \mathcal{G}(n,p)$ the critical ratio $\mathbb{E}_{\mathcal{A}}[Z^2]/\mathbb{E}_{\mathcal{A}}[Z]^2$ is polynomially bounded in $n$.
\end{proposition}
\begin{proof}
We first relate the critical ratio to the critical ratio of averages. As the estimator is unbiased $\mathbb{E}_{\mathcal{A}}[Z]=C$. Therefore, from Corollary~\ref{cor:l1},
\begin{eqnarray*} C=\mathbb{E}_{\mathcal{A}}[Z]=\frac{\mathbb{E}_{\mathcal{A}}[X]}{aut(H)} \geq   \frac{\mathbb{E}_{\mathcal{G}}[\mathbb{E}_{\mathcal{A}}[X]]}{ \omega \cdot aut(H)}. \end{eqnarray*}
Squaring both sides, \begin{eqnarray*}\mathbb{E}_{\mathcal{A}}[Z]^2=\frac{\mathbb{E}_{\mathcal{A}}[X]^2}{aut(H)^2} \geq  \frac{\mathbb{E}_{\mathcal{G}}[\mathbb{E}_{\mathcal{A}}[X]]^2}{\omega^2\,aut(H)^2}.  \end{eqnarray*}
Note that $\mathbb{E}_{\mathcal{A}}[X]/aut(H)$ refers to the expected output for fixed graph $G$, and the inequalities hold for almost all such graphs $G$, while $\mathbb{E}_{\mathcal{G}}[\mathbb{E}_{\mathcal{A}}[X]]/aut(H)$ is the expected output for a random graph $G \in \mathcal{G}(n,p)$.

The numerator of critical ratio of averages, $\mathbb{E}_{\mathcal{G}}[\mathbb{E}_{\mathcal{A}}[Z^2]]=\mathbb{E}_{\mathcal{G}}[\mathbb{E}_{\mathcal{A}}[X^2]]/aut(H)^2$. Using Markov's inequality 
\[ \Pr \left [\mathbb{E}_{\mathcal{A}}[Z^2] \geq \omega  \mathbb{E}_{\mathcal{G}}[\mathbb{E}_{\mathcal{A}}[Z^2]] \right ] \leq \frac{1}{\omega} \stackrel{n \rbec \infty}{\longrightarrow} 0 \]
Using the above inequalities yields 
\[\frac{\mathbb{E}_{\mathcal{A}}[Z^2]}{\mathbb{E}_{\mathcal{A}}[Z]^2}=\frac{\mathbb{E}_{\mathcal{A}}[X^2]}{\mathbb{E}_{\mathcal{A}}[X]^2} \leq  \omega^3 \frac{\mathbb{E}_{\mathcal{G}}[\mathbb{E}_{\mathcal{A}}[Z^2]]}{\mathbb{E}_{\mathcal{G}}[\mathbb{E}_{\mathcal{A}}[Z]]^2} = \omega^3 \frac{\mathbb{E}_{\mathcal{G}}[\mathbb{E}_{\mathcal{A}}[X^2]]}{\mathbb{E}_{\mathcal{G}}[\mathbb{E}_{\mathcal{A}}[X]]^2}.  \]

Now, we just concentrate on bounding the critical ratio of averages.  Let $V_1,\dots,V_\ell$ denote a decomposition of $H$ of width $w$. In the bipartite graph $H_i$ between the vertices of $U_i$ and $V_i$ with edge set $E_{H_i}$, let $e_i=|E_{H_i}|$, $v_i=|V_i|$, and $u_i=|U_i|$. Let $n_i=n- \sum_{j <i} v_j$. We will rely on the fact that all the $H_i$'s are of bounded (size) width. 

Let $n'_i=n_i+u_i$. Let $G_i$ be a random graph from $\mathcal{G}(n'_i,p)$ with $u_i$ distinguished vertices. Let $L_{H_i|U_i}(G_i)$ denote the number of embeddings of $H_i$ in $G_i$ where the mapping of the vertices in $U_i$ to the distinguished vertices in $G_i$ is fixed (given). The results do not depend on the mapping used for $U_i$. We abbreviate $L_{H_i|U_i}(G_i)$ by $L_i$.

First we investigate the numerator of the critical ratio of averages. Here we use the fact that
\begin{eqnarray*}
\mathbb{E}_{\mathcal{G}}[\mathbb{E}_{\mathcal{A}}[X^2]] =\mathbb{E}_{\mathcal{G}}[\mathbb{E}_{\mathcal{A}}[X_{1}^2]] \cdot \dots \cdot \mathbb{E}_{\mathcal{G}}[\mathbb{E}_{\mathcal{A}}[X_{\ell}^2]].          
\end{eqnarray*}
The previous equality arises, because at the $i$th stage the graph used for embedding $H_i$ is from $\mathcal{G}(n'_i,p)$ irrespective of the choices made over the first $(i-1)$ stages. This is guaranteed by Property~\ding{174} of the decomposition and in turn it allows us to perform a stage-by-stage analysis of the critical ratio.  

Furthermore, $\mathbb{E}_{\mathcal{G}}[\mathbb{E}_{\mathcal{A}}[X_{i}^2]]=\mathbb{E}[L_i^2]$ (as the graph is random, it doesn't matter which vertices $U_i$ gets mapped to). Next we investigate the  denominator of the critical ratio of averages. Here we use the fact that
\[\mathbb{E}_{\mathcal{G}}[\mathbb{E}_{\mathcal{A}}[X]]^2=(n!p^{e_H})^2=\mathbb{E}[L_1]^2\cdot \dots \cdot \mathbb{E}[L_{\ell}]^2.\]

Therefore, the ratio  \begin{eqnarray*} \frac{\mathbb{E}_{\mathcal{G}}[\mathbb{E}_{\mathcal{A}}[X^2]]}{\mathbb{E}_{\mathcal{G}}[\mathbb{E}_{\mathcal{A}}[X]]^2} &=&  \frac{\mathbb{E}_{\mathcal{G}}[\mathbb{E}_{\mathcal{A}}[X_{1}^2]] \cdot \dots \cdot \mathbb{E}_{\mathcal{G}}[\mathbb{E}_{\mathcal{A}}[X_{\ell}^2]]}{\mathbb{E}_{\mathcal{G}}[\mathbb{E}_{\mathcal{A}}[X_{1}]]^2 \cdot \dots \cdot \mathbb{E}_{\mathcal{G}}[\mathbb{E}_{\mathcal{A}}[X_{\ell}]]^2} \\
&=& \frac{\mathbb{E}[L_1^2]\cdot \dots \cdot \mathbb{E}[L_{\ell}^2]}{\mathbb{E}[L_1]^2\cdot \dots \cdot \mathbb{E}[L_{\ell}]^2} = \prod_{i=1}^{\ell} \frac{\mathbb{E}[L_i^2]}{\mathbb{E}[L_i]^2}  \, .  \end{eqnarray*}To bound this expression we investigate the parameter $Var[L_i]$. 

Now consider a complete bipartite graph $K_{u_i,n_i}$ with one side being the $u_i$ distinguished vertices of $G_i$ and the other side being the remaining (non-distinguished) vertices of $G_i$. Let $\mathcal{F}_{H_i|U_i}(K_{u_i,n_i})$ be the set of embeddings of $H_i$ in $K_{u_i,n_i}$ where the mapping of the vertices in $U_i$ to the distinguished vertices in $K_{u_i,n_i}$ is fixed as in $G_i$ (note that one side of $K_{u_i,n_i}$ contains these distinguished vertices) . For each embedding $f$ from $\mathcal{F}_{H_i|U_i}(K_{u_i,n_i})$ define the indicator random variable $I_{f(H_i)}=\mathbf{1}[ f(H_i) \subseteq G_i]$. For each $F \subseteq H_i$, let $e_F$ be the number of edges in $F$, and let $r_F$ be the number of vertices in $F$ which belong to $V_i$. Now there are $\Theta(n_i^{2v_i-r_F})$ pairs $(f,g)$ of embeddings of $H_i$ in $\mathcal{F}_{H_i|U_i}(K_{u_i,n_i})$ with $f(H_i) \cap g(H_i)$ isomorphic ($\simeq$) to $F$. In the following, we use $A \asymp B$ for $A=\Theta(B)$.  

\begin{eqnarray*} 
Var[L_i] & = & \sum_{f,g}Cov[I_{f(H_i)},I_{g(H_i)}] \\
& = & \sum_{\substack{f,g \\ E_{f(H_i)} \cap E_{g(H_i)} \neq \, \emptyset}} \! \mathbb{E}[I_{f(H_i)}I_{g(H_i)}]-\mathbb{E}[I_{f(H_i)}]\mathbb{E}[I_{g(H_i)}] \\
&=&\sum_{F \subseteq H_i, e_F >0} \sum_{\substack{f,g \\ f(H_i) \cap g(H_i) \simeq F}} \mathbb{E}[I_{f(H_i)}I_{g(H_i)}]-\mathbb{E}[I_{f(H_i)}]\mathbb{E}[I_{g(H_i)}] \\
& =& \sum_{F \subseteq H_i, e_F >0} \sum_{\substack{f,g \\ f(H_i) \cap g(H_i) \simeq F}} p^{2e_i-e_F}-p^{2e_i}  \\
& \asymp &   \sum_{F \subseteq H_i, e_F>0} n_i^{2v_i-r_F}(p^{2e_{i}-e_F}-p^{2e_{i}})\\
&= &\sum_{F \subseteq H_i, e_F>0}\frac{n_i^{2v_i}p^{2e_i}}{n_i^{r_F}p^{e_F}}(1-p^{e_{F}})  \\
&  \asymp &  \sum_{F \subseteq H_i, e_F>0}\frac{\mathbb{E}[L_i]^2}{n_i^{r_F}p^{e_F}} (1-p^{e_{F}}) \\
& \asymp & \max_{F \subseteq H_i, e_F >0} \frac{\mathbb{E}[L_i]^2}{n_i^{r_F}p^{e_F}} (1-p^{e_{F}}). 
\end{eqnarray*}
The second equality (above) used the fact that random variables $I_{f(H_i)}$ and $I_{g(H_i)}$ are independent if  $E_{f(H_i)} \cap E_{g(H_i)}=\emptyset$. The implicit constants in the above equivalences depend on the width of $H_i$ (a constant), but are independent of $n_i$. The quantity  
\[ \max_{F \subseteq H_i, e_{F} >0} \frac{(1-p^{e_{F}})}{n_i^{r_F}p^{e_{F}}} = O(1/n_i)  \mbox{ ($r_F=1$, provides the maximum).}  \] 
Therefore, $Var[L_i]/\mathbb{E}[L_i]^2 = O(1/n_i)$, implying $\mathbb{E}[L_i^2]/\mathbb{E}[L_i]^2 = 1+O(1/n_i)$. If $e_i=0$, then $Var[L_i]=0$, and $\mathbb{E}[L_i^2]=\mathbb{E}[L_i]^2$. Putting everything together, we obtain
\begin{eqnarray*} 
\frac{\mathbb{E}_{\mathcal{G}}[\mathbb{E}_{\mathcal{A}}[X^2]]}{\mathbb{E}_{\mathcal{G}}[\mathbb{E}_{\mathcal{A}}[X]]^2} = \prod_{i=1}^{\ell} \frac{\mathbb{E}[L_i^2]}{\mathbb{E}[L_i]^2} \leq \prod_{i=1}^{\ell} \left (1+\frac{c}{n_i} \right )= \prod_{i=1}^{\ell} \frac{n_i+c}{n_i},  
\end{eqnarray*}
for constant $c$ depending only on $w$ and $p$. Since $n=n_1>n_2> \dots>n_{\ell}$, $\prod_{i=1}^{\ell} \frac{n_i+c}{n_i}$ can be polynomially bounded (to $O(n^c)$) by a telescoping argument.
Putting everything together,  we get that the critical ratio $\mathbb{E}_{\mathcal{A}}[Z^2]/\mathbb{E}_{\mathcal{A}}[Z]^2$ is polynomially bounded in $n$. This completes the proof.
\end{proof}

Summarizing, we have the following result: if $H$ has a decomposition of bounded width $w$, then for almost all graphs $G$, running the Algorithm Embeddings $\mbox{poly}(n)\epsilon^{-2}$ times  and taking the mean of the outputs it generates results in an $(1\pm \epsilon)$-approximation for $C$. Here, $\mbox{poly}(n)$ is a polynomial in $n$ depending on $w$ and $p$. Since each run of the Algorithm Embeddings also takes polynomial time (as $H$ has bounded width decomposition), this is, an fpras. 

\begin{theorem} \label{thm:fpras}
Let $H$ be an $n$-vertex connected graph with a decomposition of width $w$ (a constant). Then,   there exists an fpras for estimating the number of copies of H in $G \in \mathcal{G}(n,p)$ for constant $p$.
\end{theorem}

\section{Graphs with Ordered Bipartite Decomposition}\label{exam}
We divide this section into subsections based on the increasing complexity of the graph classes. We will prove the following result in the remainder of this section. 

\begin{theorem}\!\footnote{The proof of this theorem follows by combining Propositions~\ref{prop:1},~\ref{prop:2},~\ref{prop:3},~\ref{prop:4}, and ~\ref{prop:5}.}  \label{thm:probg}
Let $H$ be a graph from one of the following graph classes: graphs of degree at most two, forests, bounded-length grid graphs, subdivision graphs, $\mathcal{C}_3$-free outerplanar graphs, $[\mathcal{C}_3,\mathcal{C}_5]$-free series-parallel graphs, or planar graphs of girth at least 16. Then,  there exists an ordered bipartite decomposition of $H$. Furthermore, if $H$ has bounded degree, then the decomposition has bounded width. 
\end{theorem}  

We concentrate on connected graphs $H$.\!\footnote{If $H$ is disconnected then a decomposition is obtained by combining the decomposition of all the connected components (in any order).} Let $\triangle$ be the maximum degree of any vertex in $H$. For constructing the decomposition, the following definitions are useful, 
\[  U^{i}=\bigcup_{j \leq i} U_j,\  \ V^{i} = \bigcup_{j \leq i} V_j, \  \ \mbox{and }
D^{i}=V^{i}-U^{i}. \] 
All our decomposition algorithms proceeds in steps with step $i$ creating the $(U_i,V_i)$ pair. 

\subsection{Some Easy Graph Classes} \label{simp}
We start off by considering easy graph classes such as graphs of degree at most two (paths and cycles), trees, and grid graphs. \FI{tri} illustrates some examples. 
\begin{itemize}
\item \noindent{\textbf {Paths:}} Let $H$ represent a path $(s_1,\dots,s_{k+1})$ of length $k=k(n)$. Then, the decomposition is, $V_i=\{s_i\}$ for $1 \leq i \leq k+1$. 
\begin{figure}[btp]
\centering
\hfill
\begin{minipage}[t]{.3\textwidth}
\begin{center}  
\vspace*{-.4in}
\includegraphics[width=5cm]{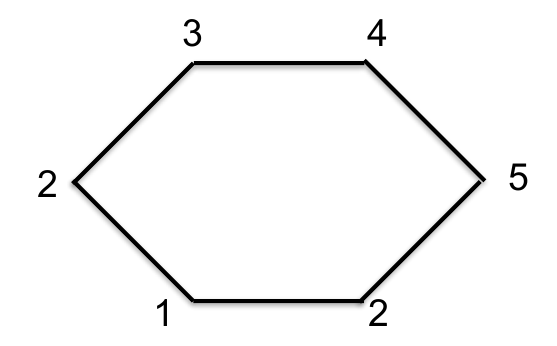}
\end{center}
\end{minipage}
\hfill
\begin{minipage}[t]{.3\textwidth}
\begin{center}  
\vspace*{-.4in}
\includegraphics[width=5cm]{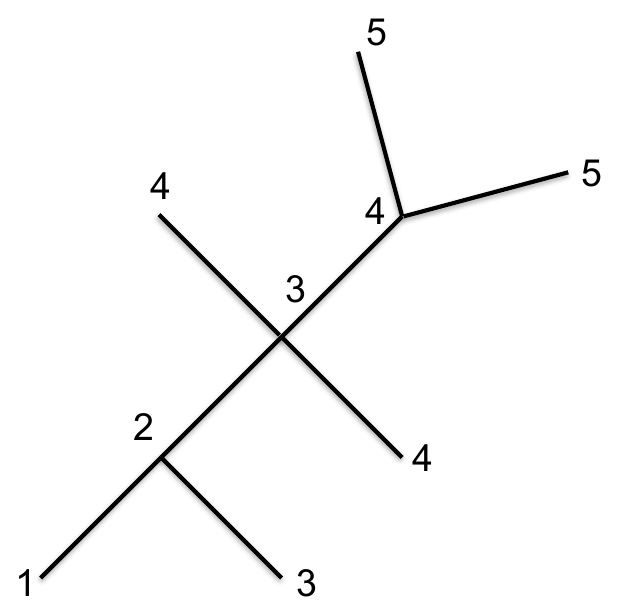}
\end{center}
\end{minipage}
\hfill
\begin{minipage}[t]{.3\textwidth}
\begin{center}  
\vspace*{-.4in}
\includegraphics[width=5cm]{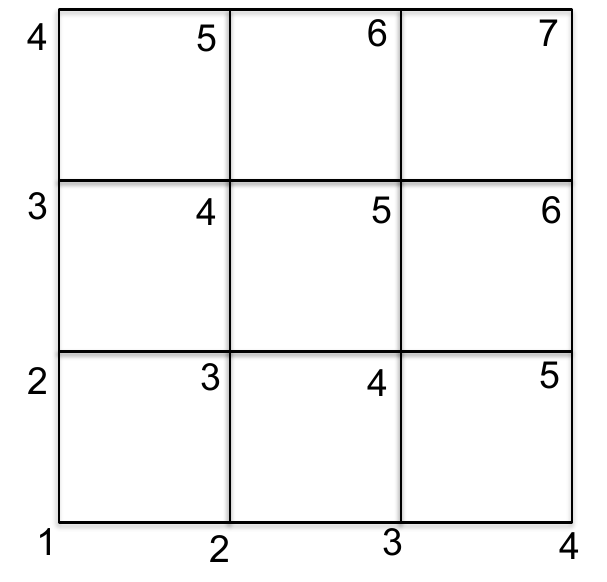}
\end{center}
\end{minipage}
\hfill
\caption{Decomposition of a cycle, tree, and grid.  Vertices with label $i$ constitute $V_i$. Neighbors of $V_i$ with lower labels constitute $U_i$.}
\label{tri}
\end{figure}

\item \noindent{\textbf {Cycles:}} First consider the cycles of length four or greater.  Let $s_1,\dots,s_k$ be the vertices of a cycle~$H$ of length $k=k(n)$ enumerated in cyclic order. In the decomposition, $V_1=\{s_1\}$, $V_2=\{s_2,s_k\}$, and $V_i=\{s_{i}\}$ for $3 \leq i \leq k-1$. Cycles of length three (triangles) don't have a decomposition, but counting copies of triangles is easy (we describe an algorithm to do so in Appendix~\ref{app:app2}). This also completes the claim for graphs of degree at most two in Theorem~\ref{thm:probg}.

\item \noindent{\textbf{Trees:}} For a tree $H$, $V_1= \{s_1\}$, where $s_1$ is any vertex in $H$. For $i \geq 2$, let $U_i$ be any vertex from $D^{i-1}$, then $V_i$ is the set of neighbors of this vertex which are not in $V^{i-1}$. Intuitively, $V_i$ is the set of children of the vertex in $U_i$, if one thinks of $H$ as a tree rooted at $s_1$. The width of this decomposition is at most $\triangle$.

\item \noindent{\textbf{Grid Graphs:}} Let  $w_0$ be the length of the grid graph $H$ (for an $n_1 \times n_2$ grid graph the length is $\min\{n_1,n_2\}$). Set $V_1=\{s_1\}$, where $s_1$ is any corner vertex in $H$. Later on, $V_i$ is the set of all vertices which are at a lattice (Manhattan) distance $i$ from $s_1$. Since for each $i$, there are at most $w_0$ vertices at distance $i$ from $s_1$, the sizes of the $V_i$'s are bounded if $w_0$ is bounded. Consequently, the width of this decomposition is bounded if $w_0$ is bounded. This construction also extends to higher dimensional grid graphs.
\end{itemize}

\begin{proposition} \label{prop:1}
Let $H$ be a graph from one of the following graph classes: graphs of degree at most two, bounded-degree forests, or bounded-length grid graphs. Then,  there exists an ordered bipartite decomposition of $H$ with bounded width.
\end{proposition}

\subsection{Decomposition of Subdivision Graphs}\label{app6}
A $k$-subdivision graph of a graph is obtained by inserting $k=k(n)$ new vertices in every edge, that is by replacing each original edge by a path of length $k+1$. We relax this definition and say that a $k$-subdivision graph is the graph obtained by inserting at least one and at most $k$ vertices in every edge. Let $H$ be a $k$-subdivision graph of a graph $F$. We now show that $H$ has a decomposition of width at most $\triangle$.

The main idea behind the decomposition is that as soon as a vertex $v$ of $F$ appears in some $V_j$, all vertices in $N_{H}(v)$ not in $V^j$ are selected in $V_{j+1}$, i.e., $v \in U_{j+1}$. The decomposition of $H$ can be formally defined as, 
\[V_i    =    \left\{
\begin{array}{l}
\{s_1\} \mbox{ where } s_1 \mbox{ is any vertex in } V_F,\mbox{ if } i=1, \\
N_{H}(a_i) - V^{i-1}, \mbox{ if } i \geq 2 \mbox{ and } \{a_i\}=V_F \cap D^{i-1} \neq \emptyset, \\
N_{H}(b_i) - V^{i-1} \mbox{ where } b_i \mbox{ is any vertex in } D^{i-1},  \mbox{ otherwise.}
\end{array} \right.  \\
\]
We now argue correctness of the decomposition for which the following lemma is useful.  
\begin{lemma} \label{sub}
There exists at most one vertex in $V_F \cap D^i$ for all $i$ in the decomposition.
\end{lemma}
\begin{proof}
Proof by induction over $i$. True by construction for $i=1$. Assume by the inductive hypothesis, $V_F \cap D^{i-1}$ has at most one vertex. If there exists a vertex in $V_F \cap D^{i-1}$, then let $a_i$ be this vertex. In this case, $N_{H}(a_i)$ doesn't contain any vertex from $V_F$ (this follows as subdivision of $F$ creates $H$). Otherwise, $b_i \notin V_F$, therefore, there is at most one vertex of $V_F$ in $V_i$ (again this follows because of subdivision of $F$ creates $H$). Therefore, in both cases, $|V_{F} \cap D^{i}| \leq 1$.  
\end{proof}

Notice that the decomposition described above selects all vertices in $H$, and the vertices selected in any $V_i$ are not selected in $V^{i-1}$, therefore, $V_i$'s form a partition of $V_H$ (Property~\ding{172}). For Property~\ding{173} notice that if  $V_i$ is constructed using $a_i$ (or $s_1$) then (by the subdivision graph construction) it is always the case that $N_{H}(a_i)$ is an independent set, and if it constructed using $b_i$, then it has at most two neighbors who do not have an edge between them (again due to the subdivision graph construction). Property~\ding{174} is satisfied as for the vertex $a_i$ or $b_i$ (or $s_1$), we select all its neighbors which are not in $V^{i-1}$ together. 

The width of this decomposition is at most  $\triangle$ as  the maximum degree of $H$ is $\triangle$. 

\begin{proposition} \label{prop:2}
Let $H$ be a subdivision of a bounded-degree graph. Then,  there exists an ordered bipartite decomposition of $H$ with bounded width.
\end{proposition}

\subsection{Decomposition of Outerplanar Graphs}
In this section, we prove the decomposition property on outer planar graphs.
A graph is outerplanar if it has a planar embedding such that all vertices are on the same face. Let $H$ be a $\mathcal{C}_3$-free outerplanar graph. The idea behind the decomposition is that vertices in $U_i$ partitions the outer face into smaller intervals, each of which can then be handled separately. 

Before we formally describe the decomposition, we need some terminology. Let $s_1,\dots,s_k$ be the vertices around the outer face with $k=k(n)$ (ordering defined by the outerplanar embedding). For symmetry, we add two dummy vertices $s_0,s_{k+1}$ without neighbors and define $U_1=\{s_0,s_{k+1}\}$, and $V_1=\{s_1\}$ (the dummy vertices play no role and can be removed before running the Algorithm Embeddings).  

The algorithm proceeds in steps with step $i$ creating the $(U_i,V_i)$ pair. For $i > 1$, two vertices $s_{j_0},s_{j_1}$ with $j_0 < j_1$, define an {\em interval} at step $i$ if $s_{j_0},s_{j_1} \in U^{i-1}$, but for $j_0 <  l < j_1, s_l \notin U^{i-1}$. If the interval is defined it is the sequence of vertices between $s_{j_0},s_{j_1}$ (including the endpoints).\!\footnote{If no interval exists, then all vertices are already part of the decomposition and we are done. Also, there could be more than one interval at each $i$, in which case, we can pick any one.}  Let $a_i$ be a median vertex of the vertices in $I \cap V^{i-1}$ (median based on the outerplanar vertex ordering), where $I$ is a step $i$ interval.  Define $U_{i}$ as the smallest subset of $V^{i-1}$ containing $\{a_i\}$ and also $N_H(N_H(U_{i})-V^{i-1}) \cap V^{i-1}$. In other words, $U_{i}$ is the smallest set of vertices in $V^{i-1}$ including $\{a_i\}$ such that the set of neighbors of $U_{i}$ excluding the vertices from $V^{i-1}$ (call this set $M_{i}$) have the property that the vertices in $M_{i}$ have no neighbors outside of $U_{i}$ in $V^{i-1}$. Note that setting $U_{i}$ as $V^{i-1}$ satisfies the above condition, but it may not be the smallest. 

Define $V_{i} = N_H(U_{i})-V^{i-1} = M_{i}$. We now  argue that this is indeed a decomposition. Consider the interval $I$ at the $i$th step, with $s_{j_0},s_{j_1}$ as the defining endpoints, and $a_i$ as the median of $I \cap V^{i-1}$.  
\begin{figure}[h]
\centering
\begin{center}  
\epsfig{file=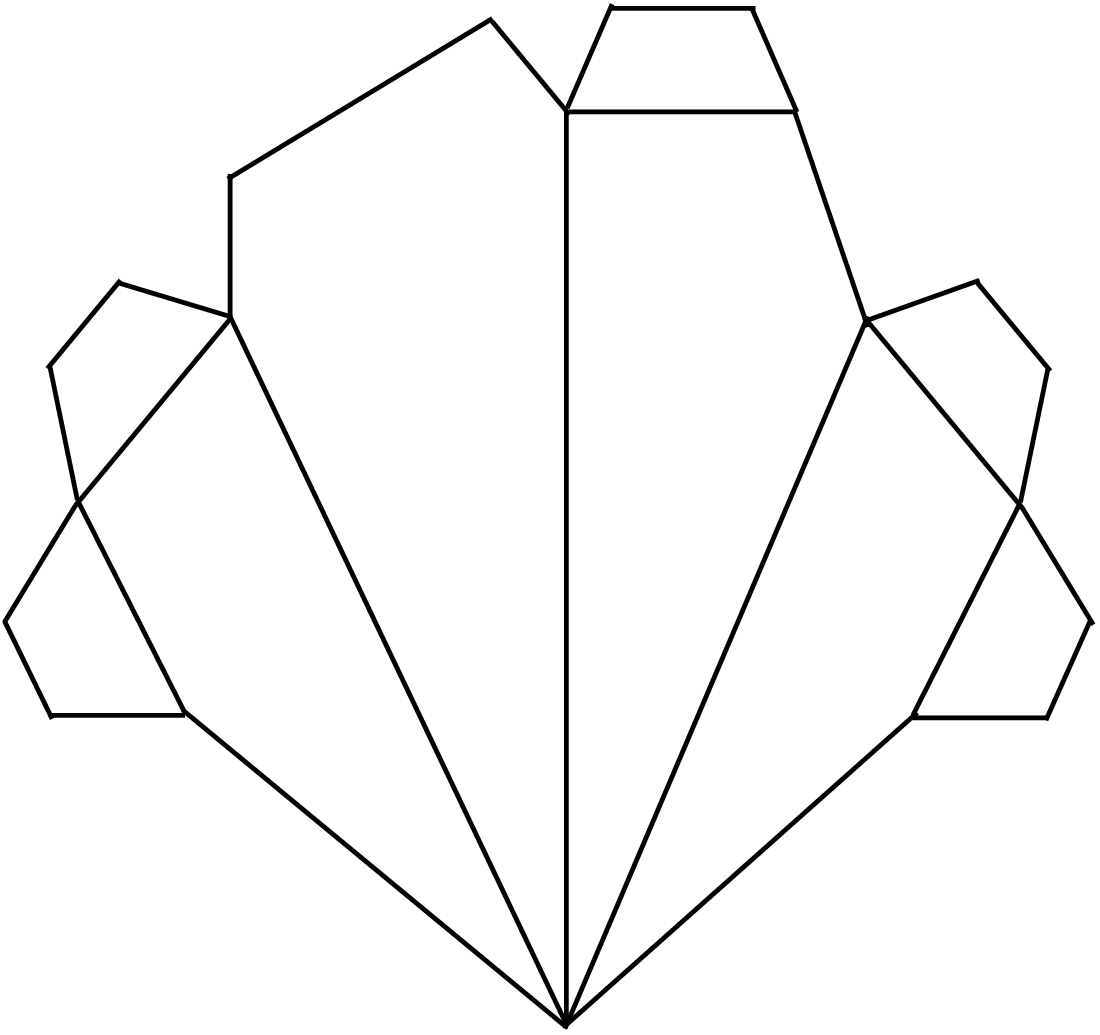, scale=0.25}
\put(-67,-7){\footnotesize 1}
\put(-27,43){\footnotesize 2}
\put(-38,80){\footnotesize 2}
\put(-62,103){\footnotesize 2}
\put(-100,80){\footnotesize 2}
\put(-120,60){\footnotesize 6}
\put(-108,40){\footnotesize 2}
\put(-4,34){\footnotesize 3}
\put(-5,78){\footnotesize 4}
\put(0,50){\footnotesize 4}                     
\put(-7,60){\footnotesize 3}
\put(-18,93){\footnotesize 3}
\put(-35,110){\footnotesize 3}
\put(-40,125){\footnotesize 5} 
\put(-60,127){\footnotesize 3}
\put(-71,123){\footnotesize 3}
\put(-112,103){\footnotesize 6} 
\put(-122,93){\footnotesize 6} 
\put(-132,83){\footnotesize 7}
\put(-137,51){\footnotesize 7}
\put(-133,32){\footnotesize 6}
\end{center}
\caption{Decomposition of an outer planar graph. Vertices with label $i$ constitute $V_i$. Neighbors of $V_i$ with lower labels constitute $U_i$.}
\label{fig:outerplanar}
\end{figure}
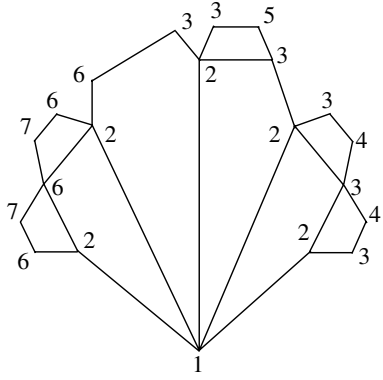

\begin{lemma} \label{fo} 
Let $I$ be the interval at the $i$th step. Then, $U_{i} \subseteq I$. 
\end{lemma}
\begin{proof}
$U_{i}$ can only contain vertices that have a path to $a_i$ but not containing any vertex from $U^{i-1}$ in the path. Since the graph is outerplanar, any path from $a_i$ to any vertex $w \notin I$ passes through either of the endpoints ($s_{j_0},s_{j_1}$), both of which are in $U^{i-1}$.  In other words, since the vertices not in $I$ do not have a path to $a_i$ which does not pass through a vertex in $U^{i-1}$, we have $U_{i} \subseteq I$. 
\end{proof}

\begin{lemma} \label{uo} 
Let $I$ be the interval at the $i$th step. Then, $|U_{i}| \leq |I \cap V^{i-1}| \leq 2\triangle$. 
\end{lemma}
\begin{proof}
The first inequality follows as $U_{i} \subseteq V^{i-1}$ (by construction) and $U_{i} \subseteq I$ (Lemma \ref{fo}).

For the second one we use induction over $i$. The hypothesis, is true by construction for $i=1$. Assume the hypothesis holds for $i-1$. Let $J$ be the interval used by the algorithm at the $(i-1)$th step. By inductive hypothesis, $|J \cap V^{i-2}| \leq 2\triangle$. The interval $J$ is split into several new intervals (at least two as $a_{i-1} \in U_{i-1}$) by the vertices of $U_{i-1}$, which define the step $i$ intervals. The newly created interval are of two types: (a) both its endpoints are from $U_{i-1}$, (b) one endpoint is from $U_{i-1}$ and other is from $U_{i-2}$. In the intervals of the first type there are at most $2\triangle$ vertices from $V_{i-1}$ (at most $\triangle$ vertices from each of the two endpoints) and no vertex from $V^{i-2}$. In the intervals of the second type, there are at most $\triangle$ vertices from $V_{i-1}$ adjacent to the endpoint in $U_{i-1}$ and at most $\triangle$ vertices from $V^{i-2}$ (from the inductive hypothesis and the fact that $a_{i-1}$ is the median of $J \cap V^{i-2}$ ). Therefore, each of the newly created step $i$ intervals (which includes $I$) have at most $2\triangle$ vertices from $V^{i-1}$. 
\end{proof}

The Properties \ding{172} and \ding{174} are guaranteed  by construction. Let us concentrate on Property~\ding{173}. For contradiction assume that there exists two vertices $v_1$ and $v_2$ in some $V_{i}$ with the edge $(v_1,v_2)$ in $H$. Since no triangles exist in $H$, both $v_1$ and $v_2$ should be connected to two different vertices (say, $u_1$ and $u_2$) in $U_{i}$. However, since the graph is outerplanar there exists no path from $u_1$ to $u_2$ going through any vertices of $U_{i} \cup V_{i}$ other than $v_1$ and $v_2$. This would mean that we could remove at least one of $u_1$ or $u_2$ from $U_{i}$ without disturbing the condition that it needs to satisfy. This would lead to a contradiction to $U_{i}$  being the smallest set in $V^{i-1}$ satisfying the condition.

Lemma \ref{uo} implies that the width of this decomposition is most $2\triangle\cdot \triangle = 2 \triangle^2 $ (as $|U_i| \leq 2 \triangle$). See \FI{fig:outerplanar} for an illustration.

\begin{proposition} \label{prop:3}
Let $H$ be a bounded-degree $\mathcal{C}_3$-free outerplanar graph. Then, there exists an ordered bipartite decomposition of $H$ with bounded width.
\end{proposition}

\subsection{Decomposition of Series-Parallel Graphs}
In this section, we prove the decomposition property on series-parallel graphs. A series-parallel graph (also called a two-terminal series-parallel graph)  is a graph with two distinguished vertices $s$ and $t$ that is obtained as follows. A single edge $(s,t)$ is a  series-parallel graph (base case).  Let $H_a$ and $H_b$ be two series-parallel graphs with terminals  $s_a,t_a$ and $s_b,t_b$ respectively. The graph formed by identifying $t_a$ with $s_b$ is a series-parallel graph with terminals $s_a,t_b$ (series operation is denoted by $\oplus$). The graph formed by identifying $s_a$ with $s_b$ and $t_a$ with $t_b$ is a series-parallel graph with terminals $s_a=s_b$ and $t_a=t_b$ (parallel operation is denoted by $||$). 

The algorithm again proceeds in steps with step $i$ creating the $(U_i,V_i)$ pair. In the following, the process of adding a vertex to some $V_i$ is referred by the term {\em selecting}. We say a vertex is {\em finished} once it is added to some $U_i$, i.e., all its neighbors are selected. The construction is technical, but the basic idea is to first finish the terminals, so that the parallel components separate (for the decomposition purposes). Then, the algorithm finishes some vertex joining two serial components. In both these steps the algorithm might be forced to finish some other vertices too. 

To define the decomposition we need some more terminology. Let $H=(V_H,E_H)$ be a $[\mathcal{C}_3,\mathcal{C}_5]$-free series-parallel graph with (distinguished) terminals $s$ and $t$.  Let $\mathcal{V}_H=V_{1,H},V_{2,H},\dots$ denote a decomposition of $H$. Let $V^i_H=\bigcup_{j \leq i} V_{j,H}$. For a set of vertices $S$ in $H$ define 
\[D_H(i,S)= \{u \in V^{i-1}_H\,:\, \mbox{ there exists } v \in S \mbox{ such that } (u,v) \in E_H\}.\] 
 $D_H(i,S)$ represents the set of neighbors of $S$ in $H$ selected in the first $(i-1)$ steps of the algorithm. The algorithm starts by finishing $s$ and $t$ as follows.  
\begin{eqnarray*} &V_{1,H}=\{s\}, \ \  \mbox{ and } \ \ V_{2,H}=N_H(s) - V^1_H.& \\
& V_{3,H}  =    \left\{
\begin{array}{l}
\{t\} \cup N_H(D_H(3,\{t\}))-V^2_H, \mbox{ if } t \notin V^2_H, \\
\emptyset, \mbox{ otherwise. }\end{array} \right.  
& \\
&V_{4,H}  = N_H(t) \cup N_H(D_H(4,N_H(t)))-V^3_H.  & 
\end{eqnarray*}  
In words, the first four steps of the algorithm achieves: (i) select $s$, (ii) finish $s$, (iii) select $t$ unless already selected, (iv) finish $t$. Define  
\[\mathcal{V}_H = V_{1,H}, V_{2,H}, V_{3,H}, V_{4,H},  \mathcal{V}_{H|s,t}\vspace{-1ex}\] where $\mathcal{V}_{H|s,t}$ is defined recursively as:

\begin{enumerate}
\item \textbf{Base case:} If all the vertices in $H$ are selected, $\mathcal{V}_{H|s,t}=\emptyset$. 
\item \textbf{Parallel case:} If $H=H_a || H_b$, find recursively $\mathcal{V}_{H_a|s,t}$ and $\mathcal{V}_{H_b|s,t}$. Define 
\[\mathcal{V}_{H|s,t} = \mathcal{V}_{H_a|s,t}, \mathcal{V}_{H_b|s,t}. \]  
\item \textbf{Serial case:} If $H=H_a \oplus H_b$, with $x$ as the vertex joining $H_a$ and $H_b$. Let  $s \in V_{H_a}$ and~$t \in V_{H_b}$.
\begin{enumerate}
\item If $x$ is finished, define $\mathcal{V}_{H|s,t} = \mathcal{V}_{H_a|s,x}, \mathcal{V}_{H_b|x,t}$. 
\item If $x \in V^4_H$ ($x$ has already been selected) and $x$ not finished, then finish $x$. This produces the set $V_{5,H}=N_H(x) \cup N_H(D_H(5,N_H(x)))-V^4_H$. Define $\mathcal{V}_{H|s,t} = V_{5,H}, \mathcal{V}_{H_a|s,x}, \mathcal{V}_{H_b|t,x}$. 
\item Otherwise, first select $x$ which produces the set $V_{5,H}=\{x\} \cup N_H(D_H(5,\{x\}))-V^4_H$. Then,  finish $x$. This produces the set $V_{6,H}=N_H(x) \cup N_H(D_H(6,N_H(x)))-V^5_H$. Define $\mathcal{V}_{H|s,t} = V_{5,H}, V_{6,H}, \mathcal{V}_{H_a|s,x}, \mathcal{V}_{H_b|t,x}$. 
\end{enumerate}
 \end{enumerate}

The following lemma provides  bounds on the sizes of $U_i$'s.  The proof looks at two possible situations, conditioning on the presence or absence of paths of length $2$ or $3$ between $s$ and $t$.  Since both $\mathcal{C}_3$ and $\mathcal{C}_5$ are forbidden, it follows that there can either be a path of length $2$ or $3$ between any two vertices, but not both. This fact will be crucial for implying Property \ding{173}. See \FI{fig:seriesparallel} for an example.

\begin{figure}[h]
\centering
\begin{minipage}[t]{.25\textwidth}
\begin{center}  
\epsfig{file=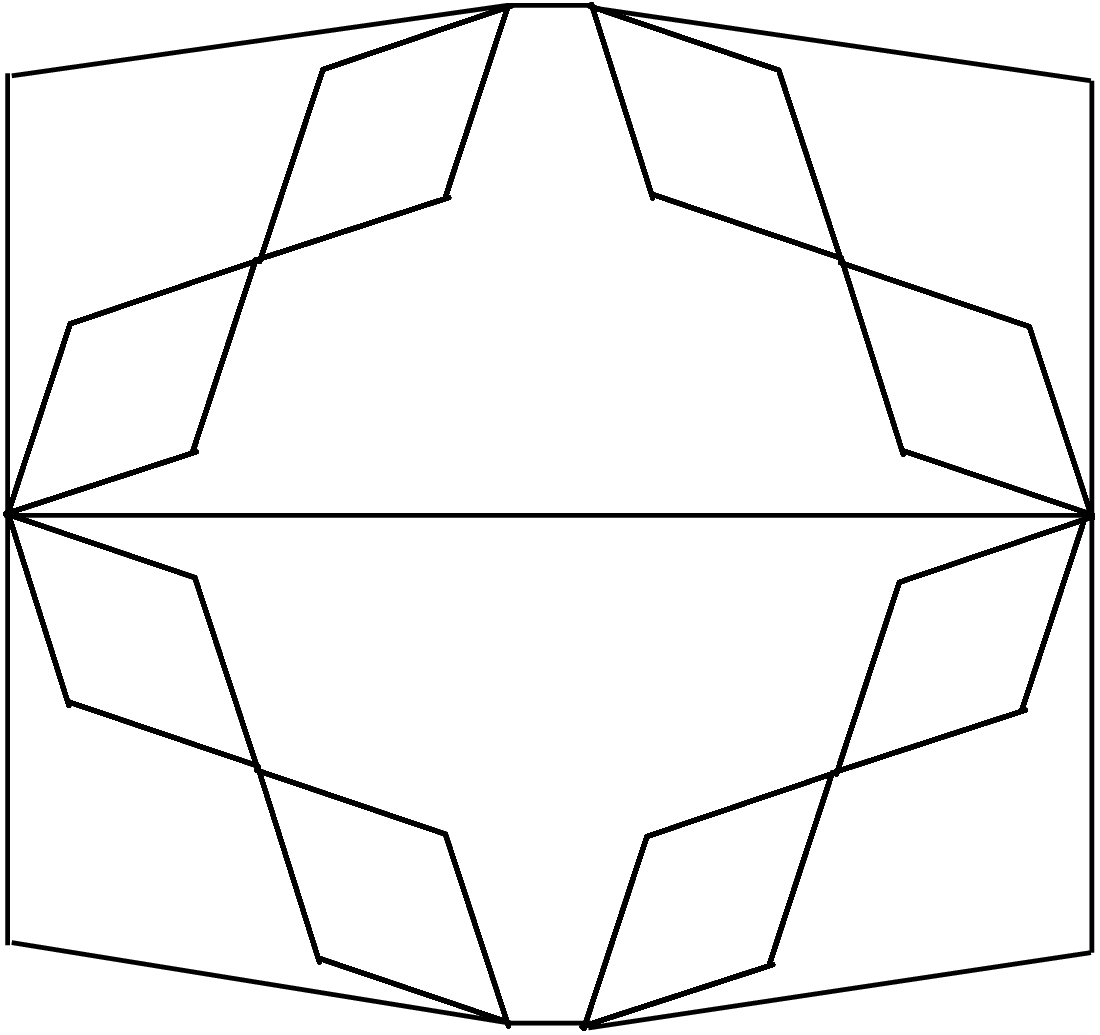, scale=0.25}
\put(-148,60){\footnotesize $1(s)$}
\put(0,60){\footnotesize $2(t)$}
\put(-74,-8){\footnotesize 9}
\put(-63,-8){\footnotesize 10}
\put(-74,129){\footnotesize 4}
\put(-61,129){\footnotesize 5}
\put(-106,8){\footnotesize 10}
\put(-36,8){\footnotesize 12}
\put(-44,33){\footnotesize 13}
\put(-80,26){\footnotesize 10}
\put(-60,26){\footnotesize 12}
\put(-100,33){\footnotesize 11}
\put(-130,37){\footnotesize 2}
\put(-106,52){\footnotesize 2}
\put(-31,52){\footnotesize 3}
\put(-106,67){\footnotesize 2}
\put(-31,67){\footnotesize 3}
\put(-7,37){\footnotesize 3}
\put(-8,86){\footnotesize 3}
\put(-130,86){\footnotesize 2}
\put(-100,87){\footnotesize 6}
\put(-37,87){\footnotesize 8}
\put(-60,92){\footnotesize 7}
\put(-35,113){\footnotesize 7}
\put(-80,92){\footnotesize 5}
\put(-101,113){\footnotesize 5}
\put(-138,118){\footnotesize 2}
\put(0,118){\footnotesize 3}
\put(-138,2){\footnotesize 2}
\put(0,2){\footnotesize 3}
\end{center}
\end{minipage}
\caption{Decomposition of a series-parallel graph. Vertices with label $i$ constitute $V_i$. Neighbors of $V_i$ with lower labels constitute $U_i$.}
\label{fig:seriesparallel}
\end{figure}
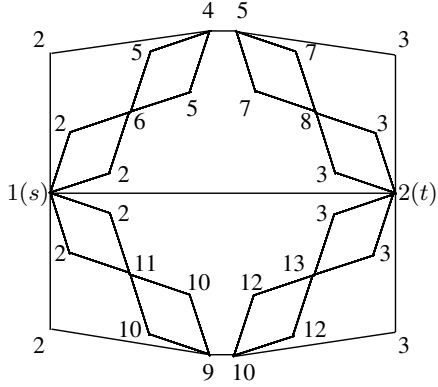

\begin{lemma}  \label{spg}
Let $H$ be a $[\mathcal{C}_3,\mathcal{C}_5]$-free series-parallel graph with terminals $s$ and $t$. Then, the above algorithm finishes $O(\triangle^2)$ vertices in every step (size of all the $U_i$'s is $O(\triangle^2)$).
\end{lemma}
\begin{proof}
The proof is via induction on the size of series-parallel graph. The inductive hypothesis is that if $s, t$ and possibly some vertices in $N_H(s) \cup N_H(N_H(s))$ are the only vertices finished, then the above algorithm finds a decomposition of $H$ by finishing $O(\triangle^2)$ vertices in every step.

The algorithm always first finishes $s$ and then $t$, and once $s$ and $t$ are finished the parallel components can be handled independently for constructing the decomposition. In the process of finishing $t$, the algorithm could possibly finish some vertices in $N_H(s) \cup N_H(N_H(s))$. Hence, in each of the parallel components $H'$, terminals $s,t$ and possibly some vertices in $N_{H'}(s) \cup N_{H'}(N_{H'}(s))$ are finished. Therefore, inductively a decomposition can be obtained. So the challenging case is when $H$ has just one parallel component. Let $H=H_1\oplus H_2$ with $z$ as the vertex joining $H_1$ and $H_2$. There are three different cases. In each of them the interesting event occurs after $s,t$, and $z$ are finished, which splits $H$ into $H_1$ and $H_2$. Afterwards, decomposition on $H_1$ and $H_2$ could be constructed independently.

In the following, we describe the cases under the assumption that there exists no edge between $s$ and $t$. If there exists such an edge, then the description would remain the same except that the step where $t$ is selected would no longer exist ($t$ is now selected when $s$ is finished). Also if there is an edge between $s$ and $t$, then there exists no path of length $2$ between $s$ and $t$, as, otherwise there would be a triangle. \\

\noindent{\textbf{Case 1: No path of length $2$ or $3$ between $s$ and $t$.}} Note that at the step when $s$ is finished no other vertex in $H$ is finished. Later, when $t$ is selected the only vertices in $N_H(s)$ that finish at that step are those which are neighbors of $t$. This set is $\emptyset$ as, otherwise, there would be a path of length $2$ between $s,t$. Similarly, at the step when $t$ is finished the only vertices in $N_H(s)$ that finish are those which share a common neighbor with $t$. This set is also $\emptyset$ as, otherwise, there would be a path of length $3$ between $s,t$. Now at the step when $z$ is selected some vertices in $N_H(s)$ and $N_H(t)$  could possibly be finished, and at the step when $z$ is finished some vertices in $N_H(s)\cup N_H(t) \cup N_H(N_H(s)) \cup N_H(N_H(t))$ could possibly be finished (this supplies the $O(\triangle^2)$ bound). However, as soon as $z$ is finished, the graphs $H_1$ and $H_2$ can be handled independently. Now $H_1$ is a smaller series-parallel graph with terminals $s,z$, where $s,z$ and possibly some vertices in $N_{H_1}(s)\cup N_{H_1}(N_{H_1}(s))$ are finished. Therefore, inductively a decomposition of $H_1$ can be completed. Similarly, $H_2$ can be viewed as a series-parallel graph with terminals $t, z$. In $H_2$, terminals $t,z$ and possibly some vertices in $N_{H_2}(t)\cup N_{H_2}(N_{H_2}(t))$ are finished. Therefore, inductively a decomposition of $H_2$ can also be completed. 
\newline

\noindent{\textbf{Case 2: Paths of length $2$ between $s$ and $t$:}} So there is no path of length $3$ between $s$ and $t$. If $t$ has been selected before $s$ is finished, then $t$ is finished together with $s$ (at which step $z$ is also selected). Note that $s$ and $t$ can be finished in the same step because there is no path of length 3. At the step when $z$ is finished some vertices in $N_H(s) \cup N_H(t)$ could possibly be finished. Afterwards,  we can invoke induction on both $H_1$ and $H_2$. If $s$ is finished before selecting $t$, then $z$ is finished while selecting $t$. At the step when $z$ is finished some vertices in $N_H(s)$ could possibly be finished. Later, at the step when $t$ is finished some vertices in $N_H(z)$ could possibly be finished. But again after $t$ is finished, we can invoke induction on both $H_1$ and $H_2$.  See \FI{fig:serpar1}.

\begin{figure}[h]
\begin{center}
\includegraphics[width=5cm]{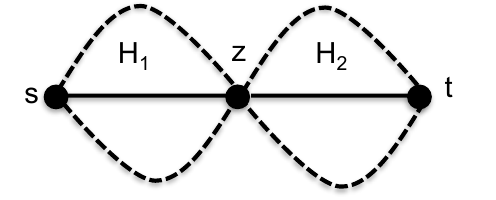}
\vspace*{-.1in}
\caption{Illustration for Case 2. The dotted edges may not be present in the graph.}
\label{fig:serpar1}
\end{center}
\end{figure}

\noindent{\textbf{Case 3: Paths of length $3$ between $s$ and $t$:}}  So there is no path of length $2$ between $s$ and $t$. There are two sub-cases based on the distance from $s$ to $z$. 

\noindent\textbf{First Sub-case:} First assume that the distance between $s$ to $z$ is one. At the step when $s$ is finished $z$ is selected. At the step when $t$ is selected no vertex in $H$ is finished (absence of path of length $2$). At the step when $t$ is finished, $z$ is  finished and also some other vertices in $N_H(s)$ could possibly be finished. Hereafter, induction can be invoked over $H_1$ and $H_2$.  See \FI{fig:serpar2}.

\begin{figure}[h]
\begin{center}
\includegraphics[width=5cm]{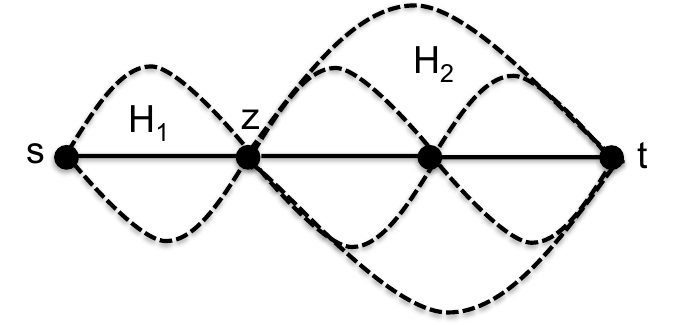}
\vspace*{-.1in}
\caption{Illustration for the first sub-case of Case 3.}
\label{fig:serpar2}
\end{center}
\end{figure} 


\begin{figure}[h]
\begin{center}
\includegraphics[width=5cm]{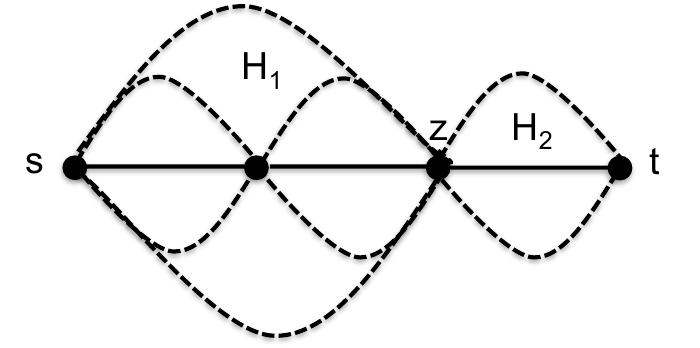}
\vspace*{-.1in}
\caption{Illustration for the second sub-case of Case 3.}
\label{fig:serpar3}
\end{center}
\end{figure} 


\noindent \textbf{Second Sub-case:} Now if the distance between $s$ and $z$ is two. Then,  the distance between $t$ and $z$ is one. At the step when $s$ is finished no other vertex in $H$ is finished. At the step when $t$ is selected no vertex in $H$ is finished. At the step when $t$ is finished, $z$ gets selected and  some vertices in $N_H(s)$ would be finished. Finally, at the step when $z$ is finished some vertices in $N_H(s) \cup N_H(N_H(s)) \cup N_H(t)$ could possibly be finished. Hereafter, induction can be invoked over $H_1$ and $H_2$.  See \FI{fig:serpar3}.

Therefore, a decomposition of $H$ can be obtained with no more than $O(\triangle^2)$ finishing at each step. A more precise upper bound of $2\triangle^2$ can be obtained by a more careful analysis. 
\end{proof}

The Properties \ding{172} and \ding{174} are guaranteed by construction.  Property~\ding{173} follows from the fact that  during any step of the above algorithm the set of vertices selected (appearing in the same $V_i$) is at most distance two (i.e, two neighborhood away) from some fixed vertex (see the proof of Lemma~\ref{spg}). Since $H$ has no $\mathcal{C}_3$ or $\mathcal{C}_5$, the vertices selected together can't have any edge between themselves (i.e., $V_i$'s are independent sets).  

The width of this decomposition is $O(\triangle^2) \cdot \triangle = O(\triangle^3)$ as $O(\triangle^2)$ vertices are finished in each step by the above algorithm (Lemma~\ref{spg}).

\begin{proposition} \label{prop:4}
Let $H$ be a bounded-degree $[\mathcal{C}_3,\mathcal{C}_5]$-free series-parallel graph. Then, there exists an ordered bipartite decomposition of $H$ with bounded width.
\end{proposition}

\subsection{Decomposition of Planar Graphs}
In this section, we prove the decomposition property on planar graphs. Define a \emph{thread} as an induced path in $H$ whose vertices are all of degree $2$ in $H$. A $k$-thread is a thread with $k$ vertices. Let $H$ be a planar graph of girth at least 16. We first prove a structural result on planar graphs. 
\begin{lemma} \label{planare}
Let $H$ be a planar graph of minimum degree $2$ and girth at least $16$, then $H$ always contains a $3$-thread. 
\end{lemma}
\begin{proof}
Assume without loss of generality that the graph $H$ is connected, otherwise it is sufficient to prove the statement for each of the components. Let $\widehat{H}$ be the graph obtained from $H$ by contracting all degree $2$ vertices. Then, $\widehat{H}$ is a planar graph of minimum degree $3$. 

We first show that $\widehat{H}$ contains a face of degree $5$ or less. For contradiction, suppose that all the faces have degree at least $6$. Let $n$ be the number of vertices, $m$ be the number of edges, and $k$ be the number of faces of $\widehat{H}$. Moreover, let $H$ be the set of faces and $V$ the set of vertices of $\widehat{H}$. Since the degree of each face is at least $6$ (where the degree of a face $f$ is the number of edges going around $f$), $2m = \sum_{f \in F} \deg(f) \geq 6k$. Moreover, $2m = \sum_{v \in V} \deg(v) \geq 3n,$ since the minimum degree in $\widehat{H}$ is at least $3$.
By Euler's formula\footnote{It states that in a planar graph with $n$ vertices,  $m$ edges, and  $k$ faces, $n-m+k=2$.}  and the previous inequalities: $m + 2 = n + k \leq (2m)/3 + m/3 = m$. A contradiction.

Let $\hat{f}$ be a face of $H$ that corresponds to a face of the degree $5$ or less in $\widehat{H}$. Since the degree of $\hat{f}$ is at least $16$ (the girth is $16$), it is easy to see that $\hat{f}$ contains a $3$-thread in $H$.
\end{proof}

In order to define a decomposition, we define a $3$-thread partition $X_1,\dots,X_c$ of a planar graph $H$ as a partition of $V_H$ such that each $X_i$ satisfies 
\[X_i   =    \left\{
\begin{array}{l}
\{a_i\}, \mbox{where } 
a_i \mbox{ is a degree $0$ or $1$ vertex in the graph induced by } V_H - \bigcup_{j < i}X_j \mbox{ on } H , 
or \\
\{a_i,b_i,c_i\}, \mbox{where } a_i,b_i,c_i \mbox{ form a $3$-thread in the graph induced by } V_H - \bigcup_{j < i}X_j \mbox{ on } H.
\end{array} \right.  \\
\]
By Lemma~\ref{planare} every planar graph with girth at least $16$ has a $3$-thread partition. As earlier, we say, a vertex is selected if we add it to some $V_k$. Using the $3$-thread partition (which can be constructed using Lemma~\ref{planare}), a decomposition of a planar graph of girth at least $16$ can be constructed by repeating this following simple procedure,
\renewcommand{\labelenumi}{\roman{enumi}.}

\begin{enumerate}
\item Find the largest index $l$ such that $X_l$ contains a vertex $z_l$ which has not yet been selected, but is adjacent to an already selected vertex. 
\item Define $U_i = N_H(z_l) \cap D^{i-1}$ and $V_i=N_{H}(U_i)-V^{i-1}$. 
\item Increment $i$. 
\end{enumerate}

\begin{lemma} \label{planarf}
Let $H$ be a planar graph of girth at least 16. Then, each of the $(U_i,V_i)$ pair created by the above algorithm satisfies that $|U_i \leq 2$ and $|V_i| \leq 2\triangle$.
\end{lemma}
\begin{proof}

Let $X_1,\dots,X_c$ be a $3$-thread partition of $H$.   Let $\bar{H}_i$ be the graph induced by  $V_H - \bigcup_{j < i}X_j$ on $H$. The first observation is that a vertex in any $X_j$ ($1 \leq j \leq c-1$) has at most one edge connecting it to the vertices in $X_{j+1} \cup \dots \cup X_c$. Consider some step $i$ of the decomposition (step $i$ is when the $(U_i,V_i)$ pair is created). Let $l$ be the largest index with an unselected vertex $z_l$.  From the previous observation it follows that vertices in $N(z_l)$ that are in $X_1 \cup \dots \cup X_{l-1}$ are not selected, in steps $1$ to $i-1$. Assume otherwise. Let $u$ be a vertex belonging to $N(z_l) \cap X_{l'} (l' < l)$ that is  selected in the first $i-1$ steps. Then, $u$ needs to have a neighbor in $X_{l} \cup \dots \cup X_{c} -\{z_l\}$, a contradiction since it would imply that $u$ (which is in $X_{l'}$) has two neighbors in $X_l \cup \dots \cup X_c$. Therefore, till step $i$ none of the neighbors of $z_l$ in $X_1 \cup \dots \cup X_{l-1}$ have been selected.  By definition of threads $z_l$ could have at most two neighbors in $\bar{H}_i$. The cases where it has two neighbors are one of the following: (a) $z_l$ has one neighbor from $X_{l+1} \cup \dots \cup X_{c}$ and another from $X_l$, or (b) $z_l$ has both its neighbors from $X_l$. This implies that $|N_H(z_l) \cap D^{i-1}|=|U_i| \leq 2$, and $|V_i| \leq 2\triangle$.
\end{proof}
The Properties \ding{172} and \ding{174} of the decomposition are again guaranteed by construction. The Property \ding{173} is satisfied because $|U_i| \leq 2$ and the vertices in $U_i$ are neighbors of $z_l$, therefore, the vertices in $V_i$ can not have edges between themselves, otherwise it will result in a cycle of length $5$. Since this holds for every $V_i$, the $V_i$'s are independent sets. 

The width of this decomposition is at most $2 \triangle$ (as $|U_i| \leq 2$ from Lemma~\ref{planarf}).

\begin{proposition} \label{prop:5}
Let $H$ be a bounded-degree planar graph of girth at least $16$. Then, there exists an ordered bipartite decomposition of $H$ with bounded width.
\end{proposition}

\section{Negative Result for Ordered Bipartite Decomposition} \label{neg}
As mentioned earlier only graphs of bounded degree have a chance of having a decomposition of bounded width. So a natural question to ask is whether all bounded-degree graphs with a decomposition have one of bounded width. In this section, we answer this question negatively by showing that every unbounded-length grid graph fails to satisfy this condition.  For simplicity, we will only consider $\sqrt{n} \times \sqrt{n}$ grid graphs, but our proof techniques extend to other cases as well. 

Let $H=(V_H,E_H)$ be a $\sqrt{n} \times \sqrt{n}$ grid graph with $V_H =\{(i,j) \,:\, 0 \leq i,j \leq \sqrt{n}-1 \}$ and $E_H=\{ ((i,j),(i',j')) \,:\, i=i' \mbox{ and } |j-j'|=1 \mbox{ or } |i-i'|=1 \mbox{ and } j=j'\}$. We now show that any decomposition of $H$ has a width of at least $\Omega(\sqrt{n})$.  Let $V_{1}, \dots,V_{\ell}$ be any decomposition of $H$.  Consider any $2 \times 2$ square of $H$ defined by vertices $a,c,b,d$ (in clockwise order).  Assume without loss of generality that the vertex $c$ has the smallest label (given by the decomposition) among vertices $a,b,c$, and $d$, and let the label on $c$ be $l$.  The two neighbors $a, b$ of the vertex $c$ always have the same label $l' > l$. The fourth vertex $d$ has any label $l''$ with $l'' \geq l$ and $l'' \neq l'$. We define a new graph $H' = (V_H,E_{H'})$ on the same set of vertices by putting the edge $(a,b)$ into $E_{H'}$.  Note that all vertices in a connected component  in $H'$ have the same label thus need to be chosen together in the decomposition (i.e., all vertices in a connected component  in $H'$  appear in the same $V_k$ in the decomposition). 

Let $\mathcal{H}_D$ be a class of graphs on vertex set $V_H$ with exactly one diagonal in every $2 \times 2$ square (and no other edges). That is any graph $H_D=(V_H,E_D)$ from $\mathcal{H}_D$ has for every $(i,j)$ with $0 \leq i,j \leq \sqrt{n}-2$ exactly one of the edges $((i,j),(i+1,j+1)), ((i,j+1),(i+1,j))$ in $E_D$ and no other edges are in $E_D$. Note that $H' \in \mathcal{H}_D$. The following theorem shows that any graph $H_D \in \mathcal{H}_D$ has the property that there is a connected component touching top and bottom or left and right (and therefore $H' \in \mathcal{H}_D$ also has this property). Note that (as mentioned before) every connected component in $H' \in \mathcal{H}_D$ would have to be chosen together in the decomposition implying that the width of the decomposition is $\Omega(\sqrt{n})$.

\begin{theorem}\label{dis}
Consider any graph $H_D \in \mathcal{H}_D$.  There exists a connected component of $H_D$ that contains at least one vertex from every row or at least one vertex from every column in the grid graph.
\end{theorem}
\begin{proof}
Assume $H_D$ does not have a connected component that contains a vertex of every row. Let $H_U=(V_U,E_U)$ be the subgraph of $H_D$ generated by all the vertices connected to the top row, i.e., $H_U$ is a collection of those connected components in $H_D$ that have at least one vertex from the top row. By assumption, $H_U$ does not contain any vertices from the bottom row. 

For every $2 \times 2$ sub-grid with vertices $a,c,b,d$ and edge $(a,b) \in E_{H'}$, we call $(a,b)$ a {\em boundary edge} if exactly one of $c,d$ is in $V_U$ and neither of $a$ or $b$ are in $V_U$. Let $H_B=(V-V_U,E_B)$ be the subgraph of $H_D$ where $E_B$ is the set of boundary edges. We assign the color red to all the vertices in $V_U$ and color black to all the vertices in $V-V_U$. Over the following two claims we make some observations about the structure of $H_B$. For a vertex $v$, let $C(v)$ indicate whether the vertex is colored red ($r$) or black ($b$).

\begin{figure}[t]
\begin{center}  
\epsfig{file=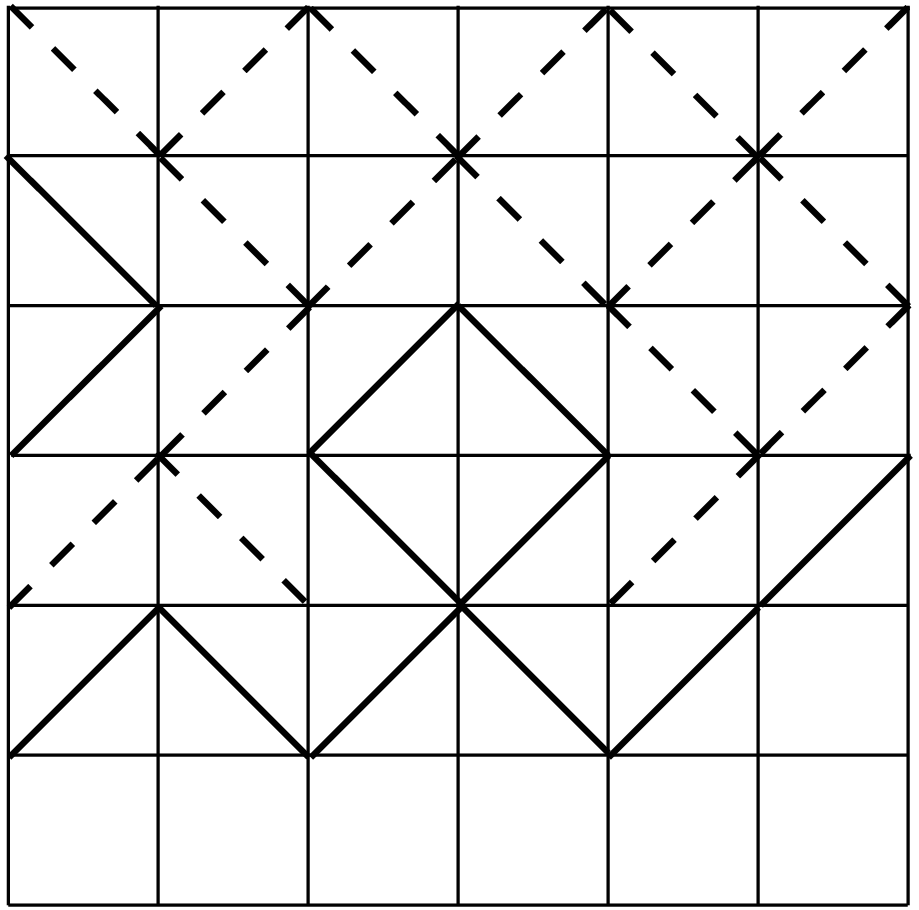, scale=.3}
\end{center}
\caption{The figure illustrates the negative result. The dotted diagonal lines are the edges in $H_U$, and the solid diagonal lines are the edges in $H_B$. There exists a component in $H_B$ than spans from the left to right boundary. } 
\label{all}
\end{figure}

\begin{claim}
There are no degree 3 vertices in $H_B$, i.e., all vertices in $H_B$ have degree $0,1,2,$ or $4$.
\end{claim}
\begin{proof}
Assume to the contrary. Let $u$ be a degree 3 black vertex. Let $(u,v_1), (u,v_2), (u,v_3)$ be the only edges incident on $u$ in $H_B$.

By choice, $C(u)=b,C(v_1)=b,C(v_2)=b,C(v_3)=b$. For the other vertices, there are only two possibilities: (i) $C(v_4)=b,C(w_1)=b,C(w_2)=r,C(w_3)=b,C(w_4)=r$ and (ii) $C(v_4)=b,C(w_1)=r,C(w_2)=b,C(w_3)=r,C(w_4)=b$.  As all the edges in $H_D$ are all either between two red vertices or two black vertices and every $2 \times 2$ sub-grid has exactly one edge, $v_4$ is black and there exists an edge between $(u,v_4)$ in $H_B$.  Therefore, every vertex in $H_B$ has degree either $0,1,2, \mbox { or } 4$. See \FI{fig:cla1}.
\end{proof}
\begin{figure}[h]
\begin{center}
\includegraphics[width=5cm]{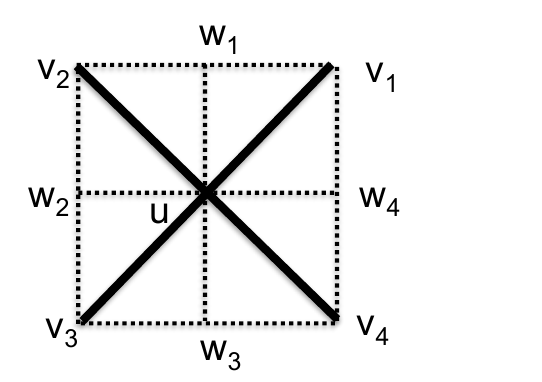}
\caption{The solid lines are the edges in $H_B$, whereas the dotted lines at the edges in the grid. If $C(u)=b,C(v_1)=b,C(v_2)=b,C(v_3)=b$, then the two possibilities of color assignments to other vertices are: (i) $C(v_4)=b,C(w_1)=b,C(w_2)=r,C(w_3)=b,C(w_4)=r$ and (ii) $C(v_4)=b,C(w_1)=r,C(w_2)=b,C(w_3)=r,C(w_4)=b$.}
\label{fig:cla1}
\end{center}
\end{figure}

As in the previous claim, by considering all possibilities for the neighbors of $u$ being in $V_U$ or not, one can conclude immediately that all vertices of degree 1 are on the left or right border and there are odd numbers of degree $1$ vertices on each border. 
\begin{claim} \label{cl2}
All the degree 1 vertices of $H_B$ are either on the left or the right border of the grid graph $H$. Additionally, there is an odd number of degree 1 vertices of $H_B$ on the left and on the right border. 
\end{claim}

Every connected component in $H_B$ has an even number of degree $1$ vertices. From Claim~\ref{cl2}, we know that degree $1$ vertices only occur at the left and right boundary of $H_B$ and there are odd number of them on both boundaries. Putting these two statements together implies that there exists a component in $H_B$ (therefore, in $H_D$) that connects the left and the right border. This finishes the proof of the Theorem~\ref{dis}. See \FI{all} for an illustration.
\end{proof}

\begin{corollary} [Negative Result]
Every decomposition of a $\sqrt{n} \times \sqrt{n}$ grid graph $H$ has a width of  $\Omega(\sqrt{n})$.
\end{corollary}

\section{Conclusions and Open Problems} \label{sec:concl}
The natural question arising from this work is what other classes of graphs have an ordered bipartite decomposition and more importantly which of them have one of bounded width. Other than the graph classes mentioned above, the bounded-degree $[\mathcal{C}_3,\mathcal{C}_5]$-free Halin graphs~\cite{hali} where degree two vertices are allowed and hexagonal grid graphs are some other interesting graph classes which have bounded width decompositions.  Most of the graph classes we considered appear to have small treewidth. So a natural question would be to relate these two decomposition schemes. However, we show in Appendix~\ref{app:treewidth} that the treewidth and the width of an ordered bipartite decomposition are incomparable.

Another interesting problem would be to investigate the general complexity of the ordered bipartite decomposition and possibly characterize its relation to other existing graph decomposition schemas. The notion of bounded width decomposition is a natural sufficient condition for the class of algorithms based on the principle of the Algorithm Count to give almost always an fpras. But the necessary condition for the general approach to work is still unclear.  Finally, a challenging open problem is to obtain any such general result for counting in arbitrary dense graphs. 

\subsection*{Acknowledgments}
We thank Andrzej Ruci\'nski for pointing us to \cite{riordan} and Piotr Berman for simplifying the proofs in Section~\ref{neg}. The authors would also like to thank Sofya Raskhodnikova, Adam Smith, and Martin Tancer for helpful comments and discussions. 

We also thank anonymous referees for pointing out an error in Definition~\ref{def:obd} in an earlier version of this paper.

\appendix
\section{Extension to the Disjoint Triangle Case} \label{app:app2}
For simplicity, we will discuss only the case where $H$ is a union of $n/3$ vertex disjoint triangles (other cases where $H$ is a union of fewer vertex disjoint triangles can be handled similarly). Even though $H$ doesn't have a decomposition, there is a simple fpras for counting copies of $H$ in random graphs. Let $s_1,\dots,s_n$ be the vertices in $H$, with every triplet $s_{3i+1},s_{3i+2},s_{3i+3}$ forming a triangle in $H$ (for $i=0,..,n/3-1$). 

Let $Z=X/aut(H)$ be the output of the Algorithm Embeddings for inputs $H$ and $G \in \mathcal{G}(n,p=\mbox{constant})$, but where each $V_i=\{s_i\}$ and $\ell =n$ (even though $V_1,\dots,V_{\ell}$ is not an ordered bipartite decomposition). As in Proposition~\ref{mainproof}, we will again investigate the ratio $\mathbb{E}_{\mathcal{G}}[\mathbb{E}_{\mathcal{A}}[X^2]]/\mathbb{E}_{\mathcal{G}}[\mathbb{E}_{\mathcal{A}}[X]]^2$ which equals the critical ratio of averages. 

The numerator, 
\begin{eqnarray*}
\mathbb{E}_{\mathcal{G}}[\mathbb{E}_{\mathcal{A}}[X^2]]&=&\mathbb{E}_{\mathcal{G}}[\mathbb{E}_{\mathcal{A}}[X_1^2X_2^2 \cdot \dots \cdot X_{n}^2]] 
=\mathbb{E}_{\mathcal{G}}[\mathbb{E}_{\mathcal{A}}[X_1^2X_2^2X_3^2]]\cdot \dots \cdot \mathbb{E}_{\mathcal{G}}[\mathbb{E}_{\mathcal{A}}[X_{n-2}^2X_{n-1}^2X_{n}^2]]. 
\end{eqnarray*}
The last equality follows because after embedding each triangle the subgraph of $G$ into which nothing has been embedded yet is random with the original edge probability $p$. Consider a representative term from this product,  
\[
\mathbb{E}_{\mathcal{G}}[\mathbb{E}_{\mathcal{A}}[X_{3i+1}^2X_{3i+2}^2X_{3i+3}^2]]=\mathbb{E}_{\mathcal{G}}[\mathbb{E}_{\mathcal{A}}[X_{3i+1}^2]]\mathbb{E}_{\mathcal{G}}[\mathbb{E}_{\mathcal{A}}[X_{3i+2}^2X_{3i+3}^2]]=(n-3i)^2\mathbb{E}_{\mathcal{G}}[\mathbb{E}_{\mathcal{A}}[X_{3i+2}^2X_{3i+3}^2]].\]
Here, as earlier, we relied on the fact the graph into which we embed the vertex $s_{3i+2}$ is random. Let $m=X_{3i+2}$ and $m'=X_{3i+3}$.  Therefore, $m$ denotes the number of ways of embedding the vertex $s_{3i+2}$ and $m'$ denotes the number of ways of embedding the vertex $s_{3i+3}$. Since the number of edges incident on the vertices in $G$ is binomially distributed, $\mathbb{E}_{\mathcal{G}}[\mathbb{E}_{\mathcal{A}}[X_{3i+1}^2X_{3i+2}^2X_{3i+3}^2]]$ equals 
\[
(n-3i)^2 \sum_{m=0}^{n-3i-1} m^2 \left ( \sum_{m'=0}^{m-1}m'^2\binom{m-1}{m'}p^{m'}(1-p)^{m-1-m'} \right ) \binom{n-3i-1}{m} p^{m}(1-p)^{n-3i-1-m}. \]

Let $L_i$ denote the number of embeddings of a triangle in a random graph from $\mathcal{G}(n-3i,p)$. Then,  the denominator  
\[\mathbb{E}_{\mathcal{G}}[\mathbb{E}_{\mathcal{A}}[X]]^2=\mathbb{E}[L_0]^2 \cdot \dots \cdot \mathbb{E}[L_{n/3-1}]^2.   \]
Note that $\mathbb{E}[L_i]=\binom{n-3i}{3}3!p^{3}$. Using the above equalities, the critical ratio of averages can be bounded to 
\begin{eqnarray*} \frac{\mathbb{E}_{\mathcal{G}}[\mathbb{E}_{\mathcal{A}}[X^2]]}{\mathbb{E}_{\mathcal{G}}[\mathbb{E}_{\mathcal{A}}[X]]^2}=\prod_{i=0}^{n/3-1}\frac{\mathbb{E}_{\mathcal{G}}[\mathbb{E}_{\mathcal{A}}[X_{3i+1}^2X_{3i+2}^2X_{3i+3}^2]]}{\mathbb{E}[L_i]^2} \leq \prod_{i=0}^{n/3-1} \left (1+\frac{c}{n-3i} \right ),  
\end{eqnarray*}
for a constant $c$. Again, we obtain a polynomial bound on the critical ratio of averages, which translates to an fpras for counting copies of $H$ in $G \in \mathcal{G}(n,p=\mbox{constant})$.

\section{Ordered Bipartite Decomposition vs.\ Treewidth} \label{app:treewidth}
In this section, we show that the treewidth and the width of an ordered bipartite decomposition are incomparable. In one direction, consider a star graph. Treewidth is 1, but no ordered bipartite decomposition of width less than $n/2$ exists. For the other direction, we consider the $1$-subdivision graph of a constant-degree expander as explained below. 

Let $H$ be a constant-degree expander graph. Consider the $1$-subdivision graph
$S(H)$ of  $H$. From Proposition~\ref{prop:2}, $S(H)$ has an ordered bipartite decomposition of bounded width. So the only fact that remains to be verified is that vertex expansion ratio of $S(H)$ is a constant.

\begin{lemma} \label{expand}
A $1$-subdivision graph of a constant-degree expander is an expander. 
\end{lemma}
\begin{proof}
Let $A$ be a set of vertices in $H$. Let $\alpha$ (= constant) denote the vertex expansion ratio of $H$ and $\triangle$ denote the maximum degree in $H$. Let $S(H)$ denote the $1$-subdivision graph of $H$.  Let $B$ be a subset of vertices from $N_{S(H)}(A)$. We consider the vertex expansion ratios for two different scenarios of $B$.  
\begin{itemize}
\item \textbf{Case $\mathbf{B=\emptyset}$}. In this case, $|N_{S(H)}(A)| \geq |N_{H}(U)| \geq \alpha |A|$. 
\item \textbf{Case $\mathbf{B \neq \emptyset}$}. First assume that, $B=N_{S(H)}(A)$. Under this assumption, $N_{S(H)}(A \cup B)= N_{H}(A)$. Say $|N_H(A)|=\lambda$. Now even if $B \subset N_{S(H)}(A)$, $|N_{S(H)}(A \cup B)|$ is at least $\lambda$. Therefore,  
\[|N_{S(H)}(A \cup B)|=\lambda \geq \alpha |A| \geq \frac{\alpha}{\triangle+1}(|A|+|B|).\]
\end{itemize} 
The final case to consider involves a set of vertices $C$ in $S(H)$, which are not in $H$. In this case, $|N_{S(H)}(C)| \geq |C|/\triangle$.

From the above case analysis it is clear that the vertex expansion ratio of $S(H)$ is a constant, and the proof follows.
\end{proof}
$S(H)$ has constant expansion which implies a treewidth $\Theta(n)$~\cite{band}, whereas $S(H)$ has an ordered bipartite decomposition of bounded width. Therefore, treewidth and the width of an ordered bipartite decomposition are incomparable.
\end{document}